\documentclass[preprintnumbers,amsmath,amssymb]{revtex4}
\usepackage{amsmath,amsfonts,latexsym,amssymb,graphicx,graphics,epsfig,subfigure,color,makeidx}
\usepackage{xcolor,diagbox}
\usepackage{multirow}

\usepackage{booktabs}
\newcommand{\papertablestyle}{%
  \small\renewcommand{\arraystretch}{1.18}%
  \setlength{\tabcolsep}{6pt}%
  \setlength{\aboverulesep}{2pt}%
  \setlength{\belowrulesep}{3pt}%
  \setlength{\cmidrulekern}{3pt}}
\newcommand{\tabletoprule}{%
  \specialrule{0.5pt}{0pt}{1.6pt}%
  \specialrule{0.5pt}{0pt}{4pt}}
\newcommand{\tablebottomrule}{%
  \specialrule{0.5pt}{4pt}{1.6pt}%
  \specialrule{0.5pt}{0pt}{0pt}}

\ifx\pdfoutput\undefined
  \DeclareGraphicsExtensions{.eps}
\else
  \ifnum\pdfoutput>0
    \DeclareGraphicsExtensions{.pdf,.png,.jpg}
  \else
    \DeclareGraphicsExtensions{.eps}
  \fi
\fi
\usepackage[colorlinks,linkcolor=blue,anchorcolor=blue,citecolor=green,urlcolor=blue]{hyperref}
\usepackage{orcidlink}
\DeclareRobustCommand{\rev}[1]{{\color{red}%
\hypersetup{linkcolor=red,citecolor=red,urlcolor=red}#1}}

\usepackage[latin1]{inputenc}

\newcommand {\be}    {\begin{equation}}
\newcommand {\ee}    {\end{equation}}
\newcommand {\beq}   {\begin{eqnarray}}
\newcommand {\eeq}   {\end{eqnarray}}

\begin{document}

\title{Quasinormal Modes of a Rotating Ho\v{r}ava Black Hole:\\ Two Perturbative Approaches to Nonseparability}

\author{Wen-Di Guo\,\orcidlink{0000-0003-2145-6587}$^{ab}$\footnote{guowd@lzu.edu.cn}}
\author{Wentao Liu\,\orcidlink{0009-0008-9257-8155}$^{ab}$\footnote{liuwentao@lzu.edu.cn}}
\author{Yu-Xiao Liu\,\orcidlink{0000-0002-4117-4176}$^{ab}$\footnote{liuyx@lzu.edu.cn, corresponding author}}

\affiliation{$^{a}$Lanzhou Center for Theoretical Physics, \\
Key Laboratory of Quantum Theory and Applications of MoE,\\
Key Laboratory of Theoretical Physics of Gansu Province,\\
Gansu Provincial Research Center for Basic Disciplines of Quantum Physics,
Lanzhou University, Lanzhou 730000, China\\
             $^{b}$Institute of Theoretical Physics $\&$ Research Center of Gravitation,\\
             School of Physical Science and Technology, \\
             Lanzhou University, Lanzhou 730000, China}

\begin{abstract}
Quasinormal modes (QNMs) provide a powerful probe of strong-field gravity. We study fundamental scalar quasinormal modes of an asymptotically
flat rotating black hole in the low-energy sector of Ho\v{r}ava gravity, whose exact scalar equation is
nonseparable due to rotation-induced Lorentz violation. We develop two complementary
approximations that both preserve the exact Kerr sector: Method I applies a
slow-rotation approximation only to the nonseparable Lorentz-violating
part, whereas Method II employs a small-$\eta$ biorthogonal spheroidal
projection, with $\eta$ the Lorentz-violating parameter. Using residual-controlled continuation and symmetric finite differences, validated by a Kerr continued-fraction benchmark and an independent
radial spectral method, we compute the linear frequency response.
Axisymmetric modes provide an exact Kerr null test. For the corotating modes studied, the real-frequency response is
positive and increases with spin. At $a/M=0.9$, the two methods reveal a roughly $2\%$ angular correction to the real response, while the imaginary response is nonmonotonic and highly sensitive to angular
projection due to internal cancellation. These results highlight
the distinct rotational signatures of Lorentz violation in black-
hole ringdown.
\end{abstract}



\maketitle

\section{Introduction}
\label{sec:introduction}

Black holes provide a setting in which the dynamical predictions of
general relativity can be confronted with observations in the
strong-field regime \cite{BertiTesting2015,Yunes2013,Wang:2025vsx,Wang:2025btn,Liu:2025wwq,LiuKRShadow2025,LiuKerrMOG2024}. Precision tests in the Solar System and in
binary pulsars have established the success of the theory over a
wide range of scales, while leaving open the question of how gravity
behaves in regimes relevant to its ultraviolet completion
\cite{Will2014,BertiTesting2015,Yunes2013}. The observation of
GW150914 opened direct access to the merger and relaxation of a
binary black hole \cite{AbbottDiscovery}. Tests based on that event
and subsequent gravitational-wave catalogs have examined waveform
consistency, propagation, and remnant properties without finding
compelling evidence for departures from general relativity
\cite{AbbottTests2016,AbbottGWTC3Tests}. Interpreting increasingly
precise measurements requires comparably controlled predictions
for black holes in alternative theories, including a clear account
of which fields propagate, which background they probe, and which
approximations enter the calculation.

For an isolated, stationary, uncharged black hole in four-dimensional
general relativity, the Kerr geometry supplies the reference
description \cite{Kerr1963}. The black-hole uniqueness results,
subject to their regularity and asymptotic assumptions, reduce
the stationary vacuum solution to a family characterized by mass
and angular momentum \cite{Israel1967,Carter1971,Robinson1975}.
The response to a perturbation contains additional information:
its quasinormal modes (QNMs) are complex-frequency resonances
selected by ingoing behavior at the horizon and outgoing behavior
at infinity. The foundations of black-hole perturbation theory
were established by Regge and Wheeler and by Zerilli
\cite{ReggeWheeler,Zerilli1970}, while early studies of scattering
and damped gravitational radiation identified the characteristic
ringing of black holes \cite{Vishveshwara1970,Press1971}.
The real and imaginary parts of a QNM frequency determine its
oscillation rate and damping time. Their role in black-hole
dynamics, their computation, and the limitations of a normal-mode
interpretation have been reviewed extensively
\cite{KokkotasReview,NollertReview,BertiReview,KonoplyaReview}.

Because the Kerr spectrum is fixed by the mass and spin, measurements
of more than one suitably identified mode can overdetermine these
parameters and test the Kerr hypothesis. This is the central idea
of black-hole spectroscopy
\cite{Echeverria1989,Dreyer2004,BertiSpectroscopy}.
Bayesian studies have quantified the conditions under which
ringdown observations can distinguish competing hypotheses
\cite{Gossan2012}, and forecasts have explored the reach of
ground- and space-based interferometers
\cite{Berti2016,Cabero2020,Carullo2018}. Coherent combination of multiple events
offers a complementary route to weaker spectral components
\cite{YangStacking}, while analyses of observed signals have
developed practical tests of the no-hair conjecture
\cite{Isi2019}. The relation between a calculated spectrum and
an observable ringdown is nevertheless nontrivial \cite{WangSpectral2026}. Mode excitation,
the onset of the perturbative regime, detector noise, and the
chosen waveform model all enter an inference. Studies of overtones
in numerical-relativity waveforms and in GW150914 illustrate both
their potential and the sensitivity of their interpretation to
analysis choices \cite{Giesler2019,Isi2019,Cotesta2022}. These issues
make it particularly important to separate a physical frequency
shift from an error introduced by an approximate perturbation
equation.

The theoretical precision of Kerr spectroscopy rests in large part
on separability. Teukolsky's formulation reduces the relevant
perturbation equations on Kerr to radial and angular problems
\cite{Teukolsky1973}; continued-fraction methods then permit accurate
calculations of their complex eigenfrequencies \cite{Leaver1985}.
The associated spheroidal eigenvalues and eigenfunctions are also
an essential part of the rotating problem \cite{BertiSpheroidal}.
Complementary approaches include barrier-based Wentzel-Kramers-Brillouin expansions
\cite{SchutzWill,IyerWill,KonoplyaWKB} and inverse-angular-momentum expansions \cite{Dolan2009}. In suitable eikonal regimes, unstable
null orbits provide a geometric interpretation of the oscillation
and damping scales \cite{CardosoGeodesics,YangGeometry,YangSelfDual2023,GuoPhotonSpheres2022,JiangDisformal2024}.
The branching of the Kerr spectrum near extremality further
illustrates why conclusions drawn from a low-spin expansion
cannot automatically be extended to rapid rotation
\cite{YangBifurcation}.

Departures from general relativity can change both the stationary
geometry and the equations governing its perturbations \cite{GuChargedKR2026}. Parametrized
ringdown frameworks organize modifications of decoupled and coupled
radial equations \cite{CardosoParam,McManus2019,YangHolonomy2024}, whereas
theory-specific and effective-field-theory analyses relate spectral
changes to the propagating fields and their couplings
\cite{Tattersall2018,Franciolini2019,LiuMOGPert2023,GuoScalarHair2023,YangHorndeskiOdd2025}. A scalar test field provides
a relatively simple probe of a modified geometry
\cite{Deng:2025uvp,Sekhmani:2025zen,Hu:2025mvm,GuoKRQNM2024,LeiRotatingHair2024,PanEMDA2007,XiaRotatingKR2026}, but its spectrum
is not generally the spectrum of gravitational radiation in the
underlying theory. In a theory with additional degrees of freedom,
even backgrounds close to Kerr can support perturbative dynamics
that cannot be inferred from the metric alone \cite{LiuDecoupling2026}. The scalar problem
is therefore useful as a controlled first step and as a means of
developing methods for non-Kerr wave equations.

Lorentz-violating gravity offers a motivated context in which to
investigate these questions. Local Lorentz symmetry is a basic
ingredient of relativistic field theory, but its status at high
energies can be examined through spontaneous symmetry breaking and
effective descriptions of preferred-frame effects
\cite{KosteleckySamuel,Colladay1998,KosteleckyGravity}.
Experimental studies and observational forecasts constrain Lorentz violation in
matter and gravity, with the interpretation depending on the
operators and sectors being tested
\cite{Mattingly2005,Liberati2013,KosteleckyTables,Joshi:2026bjl}.

Ho\v{r}ava gravity introduces a preferred foliation and anisotropic
scaling between space and time as organizing principles for a
candidate ultraviolet completion \cite{Horava2009,WangReview}.
Its low-energy formulation is related to the hypersurface-orthogonal
sector of Einstein-aether theory, in which a dynamical unit timelike
vector selects a preferred frame
\cite{JacobsonMattingly,JacobsonHorava}; the nonprojectable extension
addresses the consistency of the scalar sector
\cite{BlasExtension,BlasModels}. Preferred-frame tests, binary-pulsar
observations, and GW170817/GRB~170817A constrain the couplings and
propagation speeds
\cite{FosterPPN,YagiConstraints,YagiSensitivities,AbbottGW170817,AbbottMultimessenger,Gumrukcuoglu2018,Oost2018}.
Here we use the metric deformation parameter to characterize the
scalar spectral response; translating it into observational constraints
requires specifying the matter coupling and physical light-cone normalization.

Black holes in preferred-frame theories also expose distinctions
between metric horizons and causal boundaries for fields with
different propagation laws. Static Einstein-aether solutions
\cite{ElingBH}, the formulation of conserved charges and black-hole
mechanics \cite{FosterRadiation}, and the study of universal
horizons \cite{Berglund2012,Cropp2014} provide examples of the
additional structure that can arise. Rotation introduces further
subtleties: slowly rotating solutions in Ho\v{r}ava and
Einstein-aether gravity need not coincide, and their preferred
fields can differ qualitatively
\cite{BarausseSlow,BarausseRotating}. Spontaneous Lorentz breaking
has also motivated exact black-hole solutions in bumblebee models
\cite{Casana2018}. These constructions demonstrate that
``Lorentz-violating black hole'' does not specify a unique geometry
or perturbation problem \cite{LiuKRStatic2024,LiuExactKR2025}. For the minimally coupled relativistic
scalar considered below, we impose boundary conditions at the
outer metric Killing horizon; fields with different high-frequency
dispersion relations would require a separate treatment.

An exact rotating solution in the low-energy sector of
Ho\v{r}ava gravity was obtained by Devecio\u{g}lu and Park
\cite{Devecioglu:2024uyi}. The uncharged family studied here is asymptotically flat.
Introducing $\eta=\sqrt{\kappa\xi}-1$,
the Kerr geometry is recovered at $\eta=0$, while the nonrotating
geometry remains Schwarzschild for any $\eta$. Rotation therefore
activates the Lorentz-violating departures from Kerr, providing
a distinctive setting for investigating their effects on
strong-field wave dynamics. The horizon radii retain their Kerr
form, while the azimuthal shift and horizon angular velocity
depend on the Lorentz-violating coupling. The exact rotating
background allows these effects to be studied without first
expanding the entire metric in angular momentum. Photon propagation
and black-hole shadows have been investigated in this geometry
using numerical ray tracing \cite{Liu:2025lwj,Zhao:2025nwi,Zhao:2026wqx}, and its causal
structure has also been studied \cite{ParkChronology}.
The quasinormal spectrum provides a complementary probe of
how these rotation-induced changes in the geometry affect
oscillation frequencies and decay rates.

The main obstacle is that stationarity and axisymmetry separate
the time and azimuthal dependence, but do not guarantee separation
of the remaining radial and polar coordinates. In this geometry,
the minimally coupled scalar equation contains a Lorentz-violating
term involving both $r$ and $\theta$. Consequently, substituting
a Kerr-like product ansatz into the exact equation does not in
general yield two independent ordinary differential equations.
Systematic slow-rotation methods can handle such problems order
by order \cite{Pani2012,Pani2013,Deng:2025atg,LiuSlow2023}, but a truncation of the entire
radial operator can become inaccurate as the spin increases.
An alternative is to perturb about a separable Kerr problem in
the deformation parameter. Ghosh \textit{et al.} showed how the
leading scalar frequency correction for a weakly nonseparable
geometry can be obtained from an effective radial equation
\cite{Ghosh:2023etd}. The complex spheroidal angular problem
also calls for a suitable projection prescription; biorthogonal
and adjoint-spheroidal constructions provide a framework for
handling its non-Hermitian structure \cite{London:2020uva}.

In this paper, we develop two complementary approximations and
study the fundamental scalar spectrum of the rotating
Lorentz-violating black hole. Method I preserves the exact Kerr
background contribution to the scalar equation and applies a
slow-rotation approximation only to its nonseparable Lorentz-violating
part. Specifically, the angle-dependent kernel is replaced by the
leading term of its slow-rotation expansion, while the Kerr radial
and angular operators, the already separable frame-dragging correction,
and the algebraic dependence on $\eta$ are retained without expansion.
Its omitted angular dependence can be
diagnosed through a dimensionless geometrical factor that vanishes
at the horizon and at infinity. Method II instead expands to
first order in $\eta$ and retains the diagonal spheroidal projection
of the angle-dependent correction. The
numerical results presented here compare the two methods at common linear
order, using matched Kerr reference frequencies and extrapolated
$\eta\to0$ responses. Independent continued-fraction benchmarks and
radial, angular, and finite-difference convergence checks distinguish
numerical uncertainty from the retained angular correction.

Our analysis focuses on three related aspects of the scalar
response. First, we examine both the oscillation frequency and the
damping rate of the $(\ell,m,n)=(2,2,0)$ mode and compare the
$m=2,1,0,-1,-2$ branches. The real-frequency response grows with spin
for the corotating modes studied, whereas the imaginary response
has a different, mode-dependent behavior and can be nonmonotonic.
The exact absence of deformation terms for $m=0$ supplies an internal
consistency check. Second, we distinguish the frame-dragging contribution
from the separable or projected contribution to the radial wave
coefficient. Their partial cancellation is especially important for
the damping response. Third, we compare the two methods at the
same linear order in $\eta$, assessing their real and imaginary
responses separately. Residual and convergence checks distinguish
numerical limitations from the retained angular correction.


The paper is organized as follows. Section~\ref{sec:background}
introduces the rotating geometry, and
Sec.~\ref{sec:scalar_perturbation} derives the scalar equation and
identifies the source of nonseparability.
Sections~\ref{sec:method1} and \ref{sec:small_LV} develop Method I
and the small-$\eta$ Method II, respectively.
Section~\ref{sec:numerical_results} presents the numerical validation,
linear responses, potential decomposition, and common-order comparison.
We summarize the results and remaining limitations in
Sec.~\ref{conclusion}.

\section{Rotating black hole in Lorentz-violating gravity}
\label{sec:background}

We consider the rotating black hole solution recently obtained in the
low-energy sector of Ho\v{r}ava gravity~\cite{Devecioglu:2024uyi}.
In Boyer--Lindquist coordinates $(t,r,\theta,\phi)$, the spacetime can
be written in the form
\begin{equation}
ds^2
=
-N^2 dt^2
+\frac{\rho^2}{\Delta}dr^2
+\rho^2 d\theta^2
+\frac{\Sigma^2\sin^2\theta}{\rho^2}
\left(d\phi+N^\phi dt\right)^2 ,
\label{eq:DEHmetric}
\end{equation}
where
\begin{equation}
\begin{aligned}
\rho^2 &= r^2+a^2\cos^2\theta, \\
\Delta &= r^2-2Mr+a^2, \\
\Sigma^2 &= (r^2+a^2)\rho^2+2Mra^2\sin^2\theta ,
\end{aligned}
\label{eq:metricfunctions}
\end{equation}
and
\begin{equation}
N^2=\frac{\rho^2\Delta}{\Sigma^2},
\qquad
N^\phi
=
-\frac{2aMr\sqrt{\kappa\xi}}{\Sigma^2}.
\label{eq:lapse_shift}
\end{equation}
Here, $M$ and $a$ denote the geometric mass and rotation parameters, respectively,
while the combination $\kappa\xi$ characterizes the departure from the
general-relativistic limit. For $\kappa\xi=1$, the metric reduces to the
Kerr geometry, whereas the nonrotating limit $a=0$ gives the Schwarzschild
spacetime.

For later convenience, we introduce a dimensionless Lorentz-violating
parameter
\begin{equation}
\eta \equiv \sqrt{\kappa\xi}-1 ,
\label{eq:eta}
\end{equation}
such that the Kerr limit corresponds to $\eta=0$. Following
Ref.~\cite{Liu:2025lwj}, the line element can then be recast as
\begin{equation}
ds^2
=
ds_{\rm Kerr}^2
+
\frac{4\eta aMr\sin^2\theta}{\rho^2}
\left[
\frac{(\eta+2)aMr}{\Sigma^2}\,dt^2
-dt\,d\phi
\right].
\label{eq:metric_kerr_deformation}
\end{equation}
The distinction between geometric parameters and conserved charges matters
when comparing spectra at different couplings. In the normalization of
Ref.~\cite{Devecioglu:2024uyi}, the conserved mass and angular momentum are
\begin{equation}
 \mathcal M=16\pi\xi M,\qquad
 J=16\pi aM\sqrt{\xi/\kappa},\qquad
 \frac{J}{\mathcal M}=\frac{a}{1+\eta}.
 \label{eq:charge_map}
\end{equation}
Throughout the numerical analysis we hold the geometric $M,a$ fixed
when varying $\eta$ and use the asymptotic time coordinate of
Eq.~\eqref{eq:DEHmetric}. A comparison at fixed conserved charges is
different, since the relation between $J/\mathcal M$ and $a$ itself
depends on $\eta$.

The Kerr part of the line element is
\begin{equation}
ds_{\rm Kerr}^2
=
-\left(1-\frac{2Mr}{\rho^2}\right)dt^2
-\frac{4aMr\sin^2\theta}{\rho^2}\,dt\,d\phi
+\frac{\rho^2}{\Delta}dr^2
+\rho^2d\theta^2
+
\frac{\Sigma^2\sin^2\theta}{\rho^2}d\phi^2 .
\label{eq:kerrmetric}
\end{equation}
Equation~\eqref{eq:metric_kerr_deformation} makes the physical role of
Lorentz violation particularly transparent. The deviation from the Kerr
geometry is entirely induced by rotation and therefore vanishes when
$a=0$. Moreover, the correction affects both the $tt$ and $t\phi$
components of the metric, modifying the strong-field frame-dragging
properties of the spacetime.

Interestingly, the radial inverse-metric component is identical to that of the
Kerr spacetime,
\begin{equation}
g^{rr}=\frac{\Delta}{\rho^2},
\end{equation}
and hence the locations of the inner and outer Killing horizons are
unchanged,
\begin{equation}
r_\pm=M\pm\sqrt{M^2-a^2}.
\label{eq:horizons}
\end{equation}
Nevertheless, the angular velocity of the event horizon is modified by
the Lorentz-violating parameter. Using
$\Omega_H=-N^\phi|_{r=r_+}$, one obtains
\begin{equation}
\Omega_H
=
\frac{a(1+\eta)}{r_+^2+a^2}.
\label{eq:OmegaH}
\end{equation}
Therefore, although the horizon radii retain their Kerr values, the
rotational properties of the horizon carry explicit signatures of
Lorentz violation. These features will play an important role in the
boundary conditions and quasinormal-mode spectrum of scalar
perturbations considered below.

\section{Scalar perturbation equation and its non-separability}
\label{sec:scalar_perturbation}

We now consider a minimally coupled massless scalar field propagating
on the rotating Lorentz-violating black hole background introduced in
Sec.~\ref{sec:background}. The dynamics of the scalar field $\Phi$ is
governed by the Klein--Gordon equation
\begin{equation}
\Box\Phi
\equiv
\frac{1}{\sqrt{-g}}
\partial_\mu
\left(
\sqrt{-g}\,g^{\mu\nu}\partial_\nu\Phi
\right)
=0.
\label{eq:KG}
\end{equation}
Since the spacetime is stationary and axisymmetric, the dependence on
the coordinates $t$ and $\phi$ can be separated by writing
\begin{equation}
\Phi(t,r,\theta,\phi)
=
e^{-i\omega t+im\phi}\Psi(r,\theta),
\label{eq:scalar_ansatz}
\end{equation}
where $\omega$ is in general a complex frequency and $m$ is the
azimuthal quantum number.

For the metric given in Eq.~\eqref{eq:DEHmetric}, the determinant takes
the remarkably simple form
\begin{equation}
\sqrt{-g}=\rho^2\sin\theta,
\label{eq:metric_determinant}
\end{equation}
which is identical to that of the Kerr spacetime. The inverse metric
components relevant for the scalar perturbation equation are
\begin{equation}
g^{rr}=\frac{\Delta}{\rho^2},\qquad
g^{\theta\theta}=\frac{1}{\rho^2},\qquad
g^{tt}=-\frac{\Sigma^2}{\rho^2\Delta},\qquad
g^{t\phi}=-\frac{2aMr(1+\eta)}{\rho^2\Delta},
\label{eq:inverse_metric}
\end{equation}
together with
\begin{equation}
g^{\phi\phi}
=
g^{\phi\phi}_{\rm Kerr}
-
\frac{
4a^2M^2r^2\eta(\eta+2)
}{
\rho^2\Delta\Sigma^2
},
\label{eq:gphiphi_inverse}
\end{equation}
where
\begin{equation}
g^{\phi\phi}_{\rm Kerr}
=
\frac{\Delta-a^2\sin^2\theta}
{\rho^2\Delta\sin^2\theta}.
\end{equation}
It is worth noting that $g^{tt}$ is exactly the same as its Kerr
counterpart, whereas the Lorentz-violating corrections enter through
$g^{t\phi}$ and $g^{\phi\phi}$.

Substituting Eq.~\eqref{eq:scalar_ansatz} into
Eq.~\eqref{eq:KG} and multiplying the resulting equation by $\rho^2$,
we obtain
\begin{equation}
\left(
\mathcal{L}_{\rm K}
+
\delta\mathcal{L}_{\rm LV}
\right)
\Psi(r,\theta)=0,
\label{eq:full_scalar_equation}
\end{equation}
where $\mathcal{L}_{\rm K}$ denotes the usual scalar wave operator in
the Kerr spacetime,
\begin{equation}
\mathcal{L}_{\rm K}
=
\partial_r\left(\Delta\partial_r\right)
+
\frac{1}{\sin\theta}
\partial_\theta
\left(
\sin\theta\,\partial_\theta
\right)
+
\left[
\frac{(r^2+a^2)^2}{\Delta}\omega^2
-a^2\omega^2\sin^2\theta
-\frac{4aMr}{\Delta}\omega m
+\frac{a^2m^2}{\Delta}
-\frac{m^2}{\sin^2\theta}
\right],
\label{eq:Kerr_operator}
\end{equation}
and the modification induced by Lorentz violation is
\begin{equation}
\delta\mathcal{L}_{\rm LV}
=
-\frac{4\eta aMr}{\Delta}\omega m
+
\frac{
4\eta(\eta+2)a^2M^2r^2m^2
}{
\Delta\Sigma^2
}.
\label{eq:LV_operator}
\end{equation}

Equation~\eqref{eq:LV_operator} clearly reveals the origin of the
non-separability of the scalar perturbation equation. The first
Lorentz-violating contribution depends only on the radial coordinate
and can therefore be absorbed into the radial sector. By contrast,
the second term contains
\begin{equation}
\Sigma^2
=
(r^2+a^2)^2
-a^2\Delta\sin^2\theta,
\label{eq:Sigma_nonsep}
\end{equation}
and hence depends simultaneously on $r$ and $\theta$. Consequently,
for generic values of $a$, $\eta$, and $m$, the usual product ansatz
\begin{equation}
\Psi(r,\theta)=R(r)S(\theta)
\end{equation}
does not reduce Eq.~\eqref{eq:full_scalar_equation} to two independent
ordinary differential equations. This is in sharp contrast with the
Kerr case, where the scalar Klein--Gordon equation is completely
separable in Boyer--Lindquist coordinates. Physically, the term proportional to $\Sigma^{-2}$ encodes the fact that the Lorentz-violating frame dragging is not aligned with the Kerr horizon, causing the wave's radial propagation to be twisted by the angular geometry.

Several limiting cases provide useful consistency checks. First, in
the general-relativistic limit $\eta=0$, one has
$\delta\mathcal{L}_{\rm LV}=0$, and the standard separable Kerr
equation is recovered. Similarly, in the nonrotating limit $a=0$,
all Lorentz-violating corrections vanish and the scalar equation
reduces to that of a Schwarzschild black hole. More interestingly,
for axisymmetric scalar perturbations with
\begin{equation}
m=0,
\end{equation}
both terms in Eq.~\eqref{eq:LV_operator} vanish identically. Therefore,
within the minimally coupled scalar-field model considered here, the
$m=0$ perturbation equation coincides exactly with the corresponding
Kerr equation,
\begin{equation}
\left.
\mathcal{L}
\right|_{m=0}
=
\left.
\mathcal{L}_{\rm K}
\right|_{m=0}.
\label{eq:m0_Kerr}
\end{equation}
The Lorentz-violating effects on the scalar quasinormal spectrum are
thus intrinsically associated with non-axisymmetric modes.

The non-separable structure of
Eq.~\eqref{eq:full_scalar_equation} prevents a direct application of
the standard Kerr quasinormal-mode formalism. In the following, we
develop two complementary perturbative approaches to overcome this
difficulty. We first expand only the nonseparable Lorentz-violating
sector in $a/M$, retaining the exact Kerr spin dependence and the algebraic
dependence on the Lorentz-violating parameter. We then adopt a small
Lorentz-violation expansion around the Kerr geometry, which allows us
to retain the full dependence on the black-hole spin.

\section{Slow-rotation expansion of the nonseparable
Lorentz-violating sector}
\label{sec:method1}

We first introduce an approximation scheme in which the Kerr sector
of the scalar perturbation equation is retained exactly, while the
slow-rotation expansion is applied only to the genuinely
nonseparable Lorentz-violating contribution. This prescription differs
from the conventional slow-rotation expansion, where the entire
background spacetime is expanded around the Schwarzschild geometry.
The main advantage of the present approach is that all Kerr
spin-dependent effects are preserved, while no smallness assumption
is imposed on the Lorentz-violating parameter $\eta$.

We start directly from the exact scalar equation
\eqref{eq:full_scalar_equation} and the correction
\eqref{eq:LV_operator} derived in Sec.~\ref{sec:scalar_perturbation}.
The first term of that correction depends only on $r$ and is already
separable, so it is retained without a spin expansion. Only the
factor $\Sigma^{-2}$ in the second term couples the radial and angular
coordinates; the approximation below acts on this factor.

To isolate its angular dependence, we introduce
\begin{equation}
D(r)\equiv r^2+a^2 ,
\label{eq:method1_D}
\end{equation}
such that
\begin{equation}
\Sigma^2
=
D^2-a^2\Delta\sin^2\theta .
\label{eq:method1_Sigma}
\end{equation}
It follows that
\begin{equation}
\frac{1}{\Sigma^2}
=
\frac{1}{D^2}
\frac{1}{1-q(r,\theta)},
\label{eq:method1_Sigma_inverse}
\end{equation}
where
\begin{equation}
q(r,\theta)
\equiv
\frac{
a^2\Delta\sin^2\theta
}{
(r^2+a^2)^2
}.
\label{eq:method1_q}
\end{equation}
Outside the event horizon one has $0\leq q<1$, and therefore
Eq.~\eqref{eq:method1_Sigma_inverse} admits the expansion
\begin{equation}
\frac{1}{\Sigma^2}
=
\frac{1}{(r^2+a^2)^2}
\left[
1+q+q^2+\cdots
\right].
\label{eq:method1_Sigma_series}
\end{equation}

At leading order in the nonseparable slow-rotation correction, we
retain only the first term,
\begin{equation}
\frac{1}{\Sigma^2}
\simeq
\frac{1}{(r^2+a^2)^2}.
\label{eq:method1_main_approx}
\end{equation}
The full dependence on
$a$ contained in $\Delta$, $r^2+a^2$, the Kerr radial potential, and
the spheroidal harmonics is retained.

This approximation possesses two useful properties. At the event
horizon, where $\Delta(r_+)=0$, one has
\begin{equation}
\Sigma^2(r_+,\theta)
=
(r_+^2+a^2)^2 ,
\end{equation}
and hence Eq.~\eqref{eq:method1_main_approx} becomes exact. At spatial
infinity, it also reproduces the correct asymptotic behavior
$\Sigma^{-2}\sim r^{-4}$. The approximation error is therefore
localized mainly in the intermediate strong-field region. Moreover,
the relative error associated with the replacement
Eq.~\eqref{eq:method1_main_approx} is directly characterized by
$q(r,\theta)$.

Substituting Eq.~\eqref{eq:method1_Sigma_series} into the
Lorentz-violating contribution gives
\begin{equation}
\frac{
4\eta(\eta+2)a^2M^2r^2m^2
}{
\Delta\Sigma^2
}
=
\frac{
4\eta(\eta+2)a^2M^2r^2m^2
}{
\Delta(r^2+a^2)^2
}
+
\frac{
4\eta(\eta+2)a^4M^2r^2m^2
\sin^2\theta
}{
(r^2+a^2)^4
}
+\mathcal{O}(a^6).
\label{eq:method1_nonsep_expansion}
\end{equation}
Equation~\eqref{eq:method1_nonsep_expansion} shows explicitly that
the first genuinely angle-dependent Lorentz-violating correction
appears at $\mathcal{O}(a^4)$. Our approximation therefore neglects
the leading Lorentz-violation-induced angular mode mixing while
resumming all purely Kerr spin dependence.

Within this approximation, the perturbation equation becomes
separable. We write
\begin{equation}
\Psi(r,\theta)
=
R_{\ell m}(r)
S_{\ell m}(\theta;a\omega),
\label{eq:method1_separation}
\end{equation}
where $S_{\ell m}$ satisfies the exact scalar spheroidal equation
\begin{equation}
\frac{1}{\sin\theta}
\frac{d}{d\theta}
\left(
\sin\theta
\frac{dS_{\ell m}}{d\theta}
\right)
+
\left[
a^2\omega^2\cos^2\theta
-\frac{m^2}{\sin^2\theta}
+A_{\ell m}
\right]
S_{\ell m}
=0.
\label{eq:method1_angular}
\end{equation}
Both the spheroidal harmonic $S_{\ell m}$ and its eigenvalue
$A_{\ell m}$ are evaluated without expanding in $a\omega$.

The corresponding radial equation takes the form
\begin{equation}
0=
\frac{d}{dr}
\left(
\Delta\frac{dR_{\ell m}}{dr}
\right)
+
\Bigg[
\frac{K^2}{\Delta}
-a^2\omega^2
+2am\omega
-A_{\ell m}
-\frac{4\eta aMr\omega m}{\Delta}
+
\frac{
4\eta(\eta+2)a^2M^2r^2m^2
}{
\Delta(r^2+a^2)^2
}
\Bigg]
R_{\ell m},
\label{eq:method1_radial}
\end{equation}
where
\begin{equation}
K(r)=(r^2+a^2)\omega-am.
\label{eq:method1_KDelta}
\end{equation}
For $\eta=0$, Eq.~\eqref{eq:method1_radial} reduces exactly to the
scalar radial equation in the Kerr spacetime.

For the numerical computation of the QNM spectrum, it is convenient
to introduce the Kerr tortoise coordinate,
\begin{equation}
\frac{dr_*}{dr}
=
\frac{r^2+a^2}{\Delta},
\label{eq:method1_tortoise}
\end{equation}
and redefine the radial function according to
\begin{equation}
R_{\ell m}
=
\frac{Z_{\ell m}}
{\sqrt{r^2+a^2}} .
\label{eq:method1_radial_redefinition}
\end{equation}
The radial equation can then be cast into the Schr\"odinger-like form
\begin{equation}
\frac{d^2Z_{\ell m}}{dr_*^2}
+
V_{\ell m}^{\rm I}(r)
Z_{\ell m}
=0,
\label{eq:method1_master}
\end{equation}
with
\begin{equation}
V_{\ell m}^{\rm I}
=
V_{\ell m}^{\rm K}
+
\delta V_{\rm I}.
\label{eq:method1_total_potential}
\end{equation}
Here the exact Kerr potential is
\begin{equation}
V_{\ell m}^{\rm K}
=
\frac{
K^2-\Delta\Lambda_{\ell m}
}{
(r^2+a^2)^2
}
-
\frac{dG}{dr_*}
-G^2 ,
\label{eq:method1_Kerr_potential}
\end{equation}
where
\begin{equation}
\Lambda_{\ell m}
=
A_{\ell m}
+a^2\omega^2
-2am\omega,
\qquad
G(r)
=
\frac{r\Delta}{(r^2+a^2)^2}.
\end{equation}
We denote the Lorentz-violating correction by $\delta V_{\rm I}$,
with the mode labels suppressed, and separate its two contributions as
\begin{equation}
\delta V_{\rm I}(r)=\delta V_{\rm FD}(r)+\delta V_{\rm sep}(r).
\label{eq:method1_deltaV}
\end{equation}
Using $D=r^2+a^2$ from Eq.~\eqref{eq:method1_D}, these terms are
\begin{equation}
\delta V_{\rm FD}=-\frac{4\eta aMr\omega m}{D^2},
\qquad
\delta V_{\rm sep}=\frac{4\eta(\eta+2)a^2M^2r^2m^2}{D^4}.
\label{eq:results_fd}
\end{equation}
The subscripts FD and sep identify the frame-dragging contribution
and the separable term obtained by approximating the nonseparable
kernel, respectively; I labels Method I. The first term comes from
the radial term of Eq.~\eqref{eq:LV_operator}, multiplied by
$\Delta/D^2$ in converting to the master-wave equation. The second
comes from the replacement $\Sigma^{-2}\simeq D^{-2}$ in the same
equation. At common linear order we shall use
\begin{equation}
\delta V_{\rm sep}^{(1)}=\frac{8\eta a^2M^2r^2m^2}{D^4},
\qquad
\delta V_{\rm I}^{(1)}=\delta V_{\rm FD}+\delta V_{\rm sep}^{(1)}.
\label{eq:method1_sep_linear}
\end{equation}
Here $(1)$ means the part linear in $\eta$, including $\eta$ itself,
not first order in spin. Thus $\delta V_{\rm sep}$ retains the full
factor $\eta(\eta+2)$, whereas $\delta V_{\rm sep}^{(1)}$ keeps only
its linear part. These quantities are frequency-dependent coefficients
in the equation $Z''+VZ=0$, with the sign convention fixed by
Eq.~\eqref{eq:method1_master}.


The QNM boundary conditions are purely ingoing at the future event
horizon and purely outgoing at spatial infinity,
\begin{alignat}{2}
Z_{\ell m} &\sim e^{-i(\omega-m\Omega_H)r_*},
&\qquad r_* &\rightarrow-\infty,
\label{eq:method1_bc_horizon}
\\
Z_{\ell m} &\sim e^{+i\omega r_*},
&\qquad r_* &\rightarrow+\infty,
\label{eq:method1_bc_infinity}
\end{alignat}
where the event-horizon angular velocity is
\begin{equation}
\Omega_H
=
\frac{a(1+\eta)}
{r_+^2+a^2}.
\label{eq:method1_OmegaH}
\end{equation}
A nontrivial consistency check follows from the near-horizon limit of
the effective potential. Using
\begin{equation}
2Mr_+=r_+^2+a^2,
\end{equation}
one obtains
\begin{equation}
V_{\ell m}^{\rm I}(r_+)
=
\left(
\omega-m\Omega_H
\right)^2 ,
\label{eq:method1_horizon_check}
\end{equation}
showing that the approximate radial equation preserves the exact
ingoing-wave structure of the Lorentz-violating black-hole horizon.

Several limits provide further consistency checks. For
$\eta=0$, the exact Kerr scalar equation is recovered. For $a=0$,
all Lorentz-violating corrections vanish and the Schwarzschild
equation follows. Moreover, for axisymmetric modes with $m=0$,
\begin{equation}
\left.\delta V_{\rm I}\right|_{m=0}=0,
\end{equation}
so that their spectrum coincides with the corresponding Kerr result
within the minimally coupled scalar-field model considered here.

The kernel approximation in Method I is controlled by the magnitude
of $q(r,\theta)$. In the numerical analysis, we will therefore monitor
\begin{equation}
q_{\rm max}(a)
\equiv
\max_{r\geq r_+,\theta}
\left[
\frac{
a^2\Delta\sin^2\theta
}{
(r^2+a^2)^2
}
\right]
\label{eq:method1_qmax}
\end{equation}
as a direct measure of the neglected angular structure. Method II,
developed below, provides a complementary framework for assessing this
approximation at small Lorentz violation while retaining the full
angular dependence. Section~\ref{sec:numerical_results} compares
the numerical implementations at matched parameters and distinguishes
their operator truncations from numerical root-finding uncertainties.

\section{Quasinormal modes in the small Lorentz-violation approximation}
\label{sec:small_LV}

We now develop a complementary perturbative scheme in which the
Lorentz-violating parameter is assumed to be small,
\begin{equation}
|\eta|\ll 1,
\label{eq:small_eta}
\end{equation}
while no expansion in the black-hole spin $a/M$ is performed. This
allows us to investigate rapidly rotating black holes, provided that
the deviation from the Kerr geometry remains sufficiently small. The
metric can then be written in the perturbative form
\begin{equation}
g_{\mu\nu}
=
g_{\mu\nu}^{\rm K}
+
\eta\,h_{\mu\nu}
+
\mathcal{O}(\eta^2),
\label{eq:metric_eta_expansion}
\end{equation}
where $g_{\mu\nu}^{\rm K}$ denotes the Kerr metric. This is precisely
the class of nonseparable geometries for which the perturbative
framework of Ref.~\cite{Ghosh:2023etd} can be applied.

Expanding the Lorentz-violating contribution in
Eq.~\eqref{eq:LV_operator} to first order in $\eta$, we obtain
\begin{equation}
\delta\mathcal{L}_{\rm LV}
=
\eta\,\delta\mathcal{L}^{(1)}
+
\mathcal{O}(\eta^2),
\label{eq:LV_operator_eta_expansion}
\end{equation}
with
\begin{equation}
\delta\mathcal{L}^{(1)}
=
-\frac{4aMr\omega m}{\Delta}
+
\frac{8a^2M^2r^2m^2}{\Delta\Sigma^2}.
\label{eq:LV_operator_first_eta}
\end{equation}
Unlike the truncation of the nonseparable sector in the previous section,
the full dependence of
\begin{equation}
\Sigma^2
=
(r^2+a^2)^2-a^2\Delta\sin^2\theta
\end{equation}
is retained here. Therefore, the second term in
Eq.~\eqref{eq:LV_operator_first_eta} depends simultaneously on
$r$ and $\theta$ and the perturbation equation remains nonseparable.

Following Ref.~\cite{Ghosh:2023etd}, we use the separability of the
zeroth-order Kerr problem and expand the scalar field in terms of
scalar spheroidal harmonics,
\begin{equation}
\Phi
=
\int d\omega
\sum_{\ell,m}
\frac{Z_{\ell m}(r)}
{\sqrt{r^2+a^2}}\,
S_{\ell m}(\theta;a\omega)
e^{-i\omega t+im\phi}.
\label{eq:eta_spheroidal_expansion}
\end{equation}
The angular functions satisfy
\begin{equation}
\frac{1}{\sin\theta}
\frac{d}{d\theta}
\left(
\sin\theta\frac{dS_{\ell m}}{d\theta}
\right)
+
\left[
a^2\omega^2\cos^2\theta
-\frac{m^2}{\sin^2\theta}
+A_{\ell m}
\right]
S_{\ell m}
=0.
\label{eq:eta_angular_equation}
\end{equation}
For complex QNM frequencies, the corresponding angular operator is
in general non-Hermitian. We therefore introduce the
adjoint-spheroidal harmonics $\widetilde{S}_{\ell m}$ and choose their
normalization such that
\begin{equation}
\int d\Omega\,
\widetilde{S}_{\ell' m'}^{\,*}(\theta)
S_{\ell m}(\theta)
e^{i(m-m')\phi}
=
\delta_{\ell\ell'}\delta_{mm'}.
\label{eq:biorthogonality}
\end{equation}
The spheroidal and adjoint-spheroidal harmonics thus provide the
bi-orthogonal angular decomposition required below
\cite{London:2020uva,Ghosh:2023etd}.

The right harmonics describe the angular shape of a mode, while
the adjoint harmonics select its coefficient in the expansion.
For a non-Hermitian problem these two roles generally require
different functions. Consequently, the projection below is not
an ordinary positive average weighted by $|S_{\ell m}|^2$ and
can itself be complex.

For $\eta=0$, each $(\ell,m)$ mode obeys the standard Kerr radial
equation
\begin{equation}
\frac{d^2Z_{\ell m}}{dr_*^2}
+
V_{\ell m}^{\rm K}(r)
Z_{\ell m}
=0,
\label{eq:kerr_radial_eta}
\end{equation}
where $V_{\ell m}^{\rm K}(r)$ is given in Eq.~\eqref{eq:method1_Kerr_potential}.

The Lorentz-violating term proportional to $\Sigma^{-2}$ generates
couplings between different spheroidal multipoles with the same
azimuthal number $m$. To characterize these couplings, we define the
radially dependent angular projection
\begin{equation}
I_{\ell'\ell m}(r)
\equiv
\int d\Omega\,
\widetilde{S}_{\ell' m}^{\,*}(\theta;a\omega)
\frac{1}{\Sigma^2(r,\theta)}
S_{\ell m}(\theta;a\omega).
\label{eq:I_llm}
\end{equation}
In particular, the diagonal quantity
\begin{equation}
I_{\ell\ell m}(r)
=
\int d\Omega\,
\widetilde{S}_{\ell m}^{\,*}
\frac{1}{\Sigma^2}
S_{\ell m}
\label{eq:I_diagonal}
\end{equation}
determines the leading correction to the QNM potential, whereas
$I_{\ell'\ell m}$ with $\ell'\neq\ell$ describes the mixing between
different angular modes.

Projecting the full scalar equation onto
$\widetilde{S}_{\ell' m}$ gives, to first order in $\eta$, a coupled
system of radial equations,
\begin{equation}
\frac{d^2Z_{\ell' m}}{dr_*^2}
+
V_{\ell' m}^{\rm K}Z_{\ell' m}
+
\eta
\sum_{\ell}
\mathcal{V}_{\ell'\ell m}(r)
Z_{\ell m}
=0,
\label{eq:coupled_eta_equation}
\end{equation}
where the Lorentz-violating coupling matrix is
\begin{equation}
\mathcal{V}_{\ell'\ell m}(r)
=
\frac{
-4aMr\omega m\,\delta_{\ell'\ell}
+
8a^2M^2r^2m^2
I_{\ell'\ell m}(r)
}{
(r^2+a^2)^2
}.
\label{eq:coupling_matrix}
\end{equation}
The first contribution in Eq.~\eqref{eq:coupling_matrix} is purely
radial and therefore diagonal in $\ell$, while all nontrivial angular
mode mixing originates from the second term.

Separating the diagonal and off-diagonal contributions, we may write
Eq.~\eqref{eq:coupled_eta_equation} as
\begin{align}
\frac{d^2Z_{\ell m}}{dr_*^2}
+
\left[
V_{\ell m}^{\rm K}
+
\delta V_{\rm II}
\right]
Z_{\ell m}
=
-\eta
\sum_{\ell'\neq\ell}
\mathcal{V}_{\ell\ell'm}
Z_{\ell'm},
\label{eq:eta_diagonal_offdiagonal}
\end{align}
where the complete diagonal potential change at the retained order is
\begin{equation}
\delta V_{\rm II}(r)=\delta V_{\rm FD}(r)+\delta V_{\rm proj}(r).
\label{eq:deltaV_eta}
\end{equation}
The frame-dragging term is the same as in Eq.~\eqref{eq:results_fd},
while the projected contribution is
\begin{equation}
\delta V_{\rm proj}(r)
=\frac{8\eta a^2M^2r^2m^2}{D^2}I_{\ell\ell m}(r).
\label{eq:results_sep}
\end{equation}
The subscript proj denotes the diagonal adjoint spheroidal projection,
and II labels Method II. The angular and azimuthal mode labels are
$\ell$ and $m$; the equal first two indices in $I_{\ell\ell m}$ select
the diagonal matrix element. As for Method I, the symbols
$\delta V_{\rm FD}$, $\delta V_{\rm proj}$, and $\delta V_{\rm II}$
include $\eta$. In contrast, $\mathcal V_{\ell'\ell m}$ in
Eq.~\eqref{eq:coupled_eta_equation} is a coupling coefficient with an
explicit factor of $\eta$ outside the sum. Thus
$\delta V_{\rm II}=\eta\mathcal V_{\ell\ell m}$.
The projected term replaces, rather than supplements, the linear
separable term of Method I: the angular correction at common order
is $\delta V_{\rm proj}-\delta V_{\rm sep}^{(1)}$.

As discussed in Ref.~\cite{Ghosh:2023etd}, at first order in the
deformation parameter the off-diagonal mode couplings can be absorbed
into a redefinition of the radial eigenfunction. Introducing
\begin{equation}
Z_{\ell m}
=
X_{\ell m}
+
\eta U_{\ell m},
\label{eq:field_redefinition_eta}
\end{equation}
with $U_{\ell m}$ chosen to satisfy
\begin{equation}
\left[
\frac{d^2}{dr_*^2}
+
V_{\ell m}^{\rm K}
\right]
U_{\ell m}
=
-
\sum_{\ell'\neq\ell}
\mathcal{V}_{\ell\ell'm}
Z_{\ell'm}^{(0)},
\label{eq:U_equation_eta}
\end{equation}
where $Z_{\ell'm}^{(0)}$ denotes the corresponding Kerr solution, the
equation determining the QNM spectrum becomes
\begin{equation}
\frac{d^2X_{\ell m}}{dr_*^2}
+
V_{\ell m}^{\rm eff}(r)
X_{\ell m}
=0,
\label{eq:master_eta}
\end{equation}
with
\begin{equation}
V_{\ell m}^{\rm eff}(r)
=
V_{\ell m}^{\rm K}(r)
+
\delta V_{\rm II}(r)
+
\mathcal{O}(\eta^2).
\label{eq:effective_eta_potential}
\end{equation}
Thus, at linear order in the Lorentz-violating parameter, the QNM
frequencies can be obtained from a single one-dimensional radial
equation. The off-diagonal quantities $I_{\ell'\ell m}$ are required
for reconstructing the perturbed eigenfunctions and for quantifying
mode mixing, whereas the first-order shift of the QNM spectrum is
fully encoded in the diagonal projection $I_{\ell\ell m}$ within this
decoupling scheme.

The perturbative counting can be understood by following an
isolated Kerr mode. Its original angular channel has an amplitude
of order unity, whereas the other channels induced by the
deformation have amplitudes of order $\eta$. Their feedback into
the original channel carries another factor of $\eta$ and hence
starts at second order. This explains why the diagonal projection
suffices for the first-order frequency shift.

Accordingly, the complex QNM frequency can be expanded as
\begin{equation}
\omega_{\ell mn}
=
\omega_{\ell mn}^{\rm K}
+
\eta\,\delta\omega_{\ell mn}
+
\mathcal{O}(\eta^2),
\label{eq:qnm_eta_expansion}
\end{equation}
where $\omega_{\ell mn}^{\rm K}$ is the Kerr QNM frequency with the
same values of $(a,\ell,m,n)$. Since the correction in
Eq.~\eqref{eq:deltaV_eta} is already of first order in $\eta$, the
spheroidal harmonics, their adjoint functions, and the corresponding
angular eigenvalues entering $I_{\ell\ell m}$ may consistently be
evaluated using the zeroth-order Kerr frequency
$\omega_{\ell mn}^{\rm K}$; replacing it by the corrected frequency
would only generate terms of order $\eta^2$.

This freezing applies only to the projection in the deformation
term. The Kerr angular eigenvalue $A_{\ell m}(a\omega)$ in the
unperturbed radial operator must still be evaluated at the trial
frequency. Accordingly, the projection in $\delta V_{\rm proj}$
is evaluated as $I_{\ell\ell m}(r;\omega_{\rm K})$, where the subscript
K denotes the Kerr reference frequency of the selected mode. The
explicit frequency dependence of $\delta V_{\rm FD}$ remains evaluated
at the trial frequency in the numerical radial operator.

The QNM boundary conditions require some care. Since the horizon
locations are identical to those of the Kerr spacetime, the standard
Kerr tortoise coordinate in Eq.~\eqref{eq:method1_tortoise} can
still be employed. At spatial infinity, the Lorentz-violating
correction vanishes and the outgoing condition remains
\begin{equation}
X_{\ell m}
\sim
e^{+i\omega r_*},
\qquad
r_*\rightarrow+\infty.
\label{eq:eta_bc_infinity}
\end{equation}
At the future event horizon, however, the angular velocity is
modified according to Eq.~\eqref{eq:OmegaH}, and the ingoing
condition becomes
\begin{equation}
X_{\ell m}
\sim
e^{-i(\omega-m\Omega_H)r_*},
\qquad
r_*\rightarrow-\infty,
\label{eq:eta_bc_horizon}
\end{equation}
where
\begin{equation}
\Omega_H
=
\frac{a(1+\eta)}{r_+^2+a^2}
=
\Omega_H^{\rm K}
+
\eta\frac{a}{r_+^2+a^2}.
\label{eq:Omega_eta_expansion}
\end{equation}

The near-horizon behavior of
Eq.~\eqref{eq:effective_eta_potential} provides a useful consistency
check. Since $\Delta(r_+)=0$, one has
\begin{equation}
\Sigma^2(r_+,\theta)
=
(r_+^2+a^2)^2,
\end{equation}
which is independent of $\theta$. The bi-orthogonality relation then
immediately gives
\begin{equation}
I_{\ell\ell m}(r_+)
=
\frac{1}{(r_+^2+a^2)^2}.
\label{eq:I_horizon}
\end{equation}
It follows that
\begin{equation}
V_{\ell m}^{\rm eff}(r_+)
=
\left(
\omega-m\Omega_H
\right)^2
+
\mathcal{O}(\eta^2),
\label{eq:potential_horizon_check}
\end{equation}
showing that the effective radial equation reproduces precisely the
expected ingoing-wave behavior at the event horizon.

Several additional limits are immediate. For $\eta=0$,
Eq.~\eqref{eq:master_eta} reduces to the exact Kerr scalar equation.
For $a=0$, the entire correction vanishes and the Schwarzschild
spectrum is recovered. Furthermore, since
Eq.~\eqref{eq:deltaV_eta} is proportional to either $m$ or $m^2$,
\begin{equation}
\left.\delta V_{\rm II}\right|_{m=0}=0,
\end{equation}
and hence
\begin{equation}
\delta\omega_{\ell 0 n}=0.
\label{eq:m0_eta_result}
\end{equation}
Therefore, within the minimally coupled scalar-field model considered
here, only non-axisymmetric modes are sensitive to Lorentz violation.

The small-$\eta$ approach, hereafter Method II, is complementary to
Method I in Sec.~\ref{sec:method1}. Both retain the exact Kerr sector,
but they truncate different parts of the Lorentz-violating correction.
Method I retains the factor $\eta(\eta+2)$ while dropping the angular
dependence encoded in $q$, whereas Method II retains that angular
dependence at first order in $\eta$. Agreement is therefore expected
in their common perturbative regime. The numerical comparison below
uses the diagonal angular projection in Eq.~\eqref{eq:deltaV_eta}
frozen at the Kerr frequency.

\section{Numerical results and physical interpretation}
\label{sec:numerical_results}

\subsection{Numerical implementation}

We consider scalar modes with $\ell=2$, $n=0$, and
$\chi=a/M=0.1,\ldots,0.9$. We write
$\widehat\omega=M\omega$ and
$\gamma=-\operatorname{Im}\widehat\omega$, so that positive
$\gamma$ denotes decay. The Kerr baseline always has the same
$(\ell,m,n)$ and spin as the deformed mode. The radial equation
is solved by matching ingoing and outgoing logarithmic derivatives.
With our $e^{-i\omega t}$ convention, the amplitude decays as
$e^{\operatorname{Im}\omega\,t}$; a more negative imaginary part
therefore means faster damping. The labels $(\ell,m,n)$ identify
the Kerr branch followed continuously as the deformation is turned
on, rather than implying exact separability of the deformed problem.

\emph{Angular projection.} We first obtain a validated Kerr
frequency for the chosen branch. The spheroidal and adjoint
functions are represented in a spherical-harmonic basis with
$|m|\leq j\leq j_{\max}$, and their projection integral is evaluated
using $N_G$ Gauss--Legendre nodes in $u=\cos\theta$~\cite{GolubWelsch1969}.
Increasing $j_{\max}$ improves the representation of a single spheroidal
mode.
For Method II the adjoint spheroidal projection is frozen at the
validated Kerr frequency, whereas the Kerr angular eigenvalue is
updated at each trial frequency.

\emph{Radial matching.} Let $Z$ denote the radial master variable
in either method, with $Z=X_{\ell m}$ for Method II. Boundary
series supply initial data at two finite radii: $r=r_++\epsilon$,
just outside the horizon, and $r=r_\infty$, where $r_\infty$ denotes
a finite outer cutoff rather than spatial infinity. The outgoing
asymptotic series is evaluated at this outer cutoff. We integrate
the inner solution outward from $r_++\epsilon$ and the outer
solution inward from $r_\infty$ to $r_{\rm match}$.
The known ingoing or outgoing wave factor is removed before
integrating the logarithmic derivative of the remaining amplitude;
it is restored in the matching condition. Writing $Y=d\ln Z/dr$,
we seek a zero of the dimensionless mismatch
\begin{equation}
F(\widehat\omega;\chi,\eta)
=M\cdot\left[Y_H(r_{\rm match})-Y_\infty(r_{\rm match})\right].
\label{eq:numerical_matching_residual}
\end{equation}
The prefactor $M$ is the black-hole mass parameter and makes $F$
dimensionless. The subscripts $H$ and $\infty$ label the solutions
initialized with horizon-ingoing and asymptotically outgoing boundary
data, respectively; both logarithmic derivatives are evaluated at
the finite matching radius $r_{\rm match}$.
Equality of these logarithmic derivatives means that the two
solutions can be joined by an overall normalization. Thus the
frequency is selected by compatibility of the two boundary
conditions.
For Method II the horizon wave-number branch is chosen
continuously from Kerr using the truncated operator's horizon
coefficient, as discussed below in the finite-deformation analysis.

\emph{Frequency search.} All integrations use $M=1$. A trial
complex frequency is corrected by the Newton step
\begin{equation}
\delta\widehat\omega_{\mathrm N}
=-\frac{F}{\partial F/\partial\widehat\omega},
\label{eq:numerical_newton_step}
\end{equation}
where the derivative is evaluated by a centered frequency
difference. Large steps are limited, and a line search reduces a
step if the residual would increase. Starting from Kerr, accepted
roots are continued to the desired positive or negative $\eta$.
Finally, paired roots at opposite deformations determine the
principal observable,
\begin{equation}
C_{\ell m}(\chi)=
\left.\frac{\partial\widehat\omega_{\ell m0}}{\partial\eta}\right|_{\eta=0},
\qquad
\widehat\omega=\widehat\omega_{\rm K}+\eta C_{\ell m}+O(\eta^2).
\label{eq:results_linear_response}
\end{equation}
Thus the principal comparison probes the common linear order.
In particular, $\operatorname{Re}C$ is the initial change of the
oscillation frequency at fixed spin, whereas
$\partial_\eta\gamma|_0=-\operatorname{Im}C$ is the initial
change of the damping rate.

The response $C_{\ell m}$ in Eq.~\eqref{eq:results_linear_response}
is dimensionless and equals $M\delta\omega_{\ell m0}$ in
Eq.~\eqref{eq:qnm_eta_expansion}; it measures the initial complex
frequency shift per unit deformation at fixed geometric $M,a$.
For comparison with a fixed-conserved-charge prescription, let
$C_\chi\equiv C_{\ell m}(\chi)$ and define $C_j$ as the derivative
of $M\omega$ with respect to $\eta$ at fixed $\mathcal M,J,\xi$,
evaluated at $\eta=0$ on the same mode branch. With $\xi$ fixed,
Eq.~\eqref{eq:charge_map} gives a constant conversion factor
$\mathcal G=(16\pi\xi)^{-1}$ and a charge-normalized spin
$j=J/(\mathcal G\mathcal M^2)=\chi/(1+\eta)$.
Fixing $\mathcal M$ fixes $M$, but fixing $j$ makes the geometric
spin vary as $\chi=(1+\eta)j$. The chain rule therefore gives
\begin{equation}
 C_j=C_\chi+\chi\frac{d\widehat\omega_{\rm K}}{d\chi}.
 \label{eq:fixed_charge_response}
\end{equation}
The second term is the change in the Kerr frequency caused by this
change of spin; it is not an additional Lorentz-violating term in the
wave operator. Consequently, the exact $m=0$ null response applies
to $C_\chi$, not generally to $C_j$. The factor $\mathcal G$ specifies
the charge normalization here; identifying it with a measured Newton
constant would require specifying the matter sector.

\subsection{Numerical accuracy and convergence}

The safeguarded Newton solver permits up to 40 updates and accepts
a root only when both the dimensionless matching residual and the
final computed Newton correction are below $10^{-12}$.
The frequency, residual, iteration count, last correction, and
convergence flag are retained for every solve. Continuation in
$\eta$ uses only an accepted root; a failed parameter step is
bisected and retried. A failed root is excluded from physical
tables and fits. Kerr seeds are additionally checked against the
continued-fraction equation, with adaptive spin continuation when
the initial root search fails.

The two stopping conditions test different aspects of the solve:
$|F|$ measures the boundary mismatch, whereas the full Newton
correction estimates the remaining frequency displacement locally.
The latter is tested before line-search damping, since a small
imposed step alone would not demonstrate convergence. Roots passing both tests are further checked against changes in
the numerical settings.

The reference settings are 24th-order boundary series,
$r_\infty=70M$, an inner offset $10^{-5}M$,
$r_{\rm match}=5M$, and 80-digit working precision. At
$\chi=0.1,0.5,0.9$ and $\eta=\pm0.01$, we independently
vary the series order through $12,16,20,24$, $r_\infty/M$ through
$40,50,70,100$, the inner offset $\epsilon/M$ through $10^{-4},10^{-5},10^{-6}$,
$r_{\rm match}/M$ through $4,5,6$, and the working precision
through $40,60,80$. Table~\ref{tab:numerical_convergence} reports
the maximum componentwise variations, including the final Newton
corrections. Values below $10^{-12}$ are reported as upper limits.

Varying the series order and the two endpoint locations tests how
well the finite starting data represent the asymptotic boundary
conditions. Moving the matching point tests the consistency of
the two integrations, while changing the working precision probes
their numerical sensitivity.

\begin{table}[t]
\centering
\caption{Radial convergence for Method II, $(\ell,m,n)=(2,2,0)$. Each error column is the maximum of the real- and imaginary-component variations over the specified sweep relative to the reference setting, including final Newton corrections. The electronic table retains both components and the number of settings. Values below the root acceptance scale are given as upper limits.}
\label{tab:numerical_convergence}
\papertablestyle
\footnotesize
\setlength{\tabcolsep}{4.5pt}
\begin{tabular}{rrrrrrrrr}
\tabletoprule
\multicolumn{2}{c}{Parameters} & \multicolumn{1}{c}{Reference frequency} & \multicolumn{5}{c}{Maximum frequency variation} & \multicolumn{1}{c}{Residual} \\
\cmidrule[0.4pt](lr){1-2}\cmidrule[0.4pt](lr){4-8}
\multicolumn{1}{c}{$\chi$} & \multicolumn{1}{c}{$\eta$} & \multicolumn{1}{c}{$\widehat\omega$} & \multicolumn{1}{c}{Series} & \multicolumn{1}{c}{$r_\infty$} & \multicolumn{1}{c}{Offset} & \multicolumn{1}{c}{Match} & \multicolumn{1}{c}{Precision} & \multicolumn{1}{c}{$|F|$} \\
\midrule[0.45pt]
0.1 & $+0.01$ & $0.4996475449-0.0966659968i$ & $5.81\!\times\!10^{-11}$ & $<10^{-12}$ & $<10^{-12}$ & $<10^{-12}$ & $9.08\!\times\!10^{-12}$ & $1.54\!\times\!10^{-21}$\\
0.1 & $-0.01$ & $0.4993170560-0.0966659967i$ & $5.58\!\times\!10^{-11}$ & $<10^{-12}$ & $<10^{-12}$ & $<10^{-12}$ & $8.86\!\times\!10^{-12}$ & $1.52\!\times\!10^{-21}$\\
\addlinespace[3pt]
0.5 & $+0.01$ & $0.5872739735-0.0935021728i$ & $8.20\!\times\!10^{-12}$ & $<10^{-12}$ & $<10^{-12}$ & $<10^{-12}$ & $7.13\!\times\!10^{-12}$ & $9.51\!\times\!10^{-14}$\\
0.5 & $-0.01$ & $0.5847142810-0.0934847908i$ & $9.58\!\times\!10^{-12}$ & $<10^{-12}$ & $<10^{-12}$ & $<10^{-12}$ & $6.06\!\times\!10^{-12}$ & $8.16\!\times\!10^{-14}$\\
\addlinespace[3pt]
0.9 & $+0.01$ & $0.7863501236-0.0692848181i$ & $<10^{-12}$ & $<10^{-12}$ & $<10^{-12}$ & $<10^{-12}$ & $<10^{-12}$ & $1.15\!\times\!10^{-15}$\\
0.9 & $-0.01$ & $0.7770076029-0.0692518709i$ & $<10^{-12}$ & $<10^{-12}$ & $<10^{-12}$ & $<10^{-12}$ & $<10^{-12}$ & $2.83\!\times\!10^{-16}$\\
\tablebottomrule
\end{tabular}
\end{table}

At $\eta=0$, an independent Leaver continued-fraction calculation
provides a benchmark. It determines the spectrum from a series
recurrence rather than matching two numerical integrations, and
therefore checks a different representation of the Kerr problem.
The matching roots in
Table~\ref{tab:kerr_benchmark} are obtained from deliberately
displaced initial guesses. The largest relative difference is
$4.1\times10^{-11}$. For the
angular projection, all combinations of
$j_{\max}=17,21,25$ and $N_G=64,96,128$ are compared at
$\chi=0.5,0.9$ on 401 exterior radial points. Relative to the
reference setting $(j_{\max},N_G)=(25,128)$, namely a spherical-basis
cutoff $j_{\max}=25$ and 128 angular quadrature nodes, the maximum
relative changes in $I_{222}$ are
$1.43\times10^{-51}$ and $2.35\times10^{-41}$ in 80-digit
arithmetic. Thus the smaller setting $(j_{\max},N_G)=(17,64)$ is adequate
at the accuracy relevant here; radial and finite-difference errors dominate.
\begin{table}[t]
\centering
\caption{Independent scalar Kerr benchmark for $(2,2,0)$. Matching is started from a displaced continued-fraction seed. Relative errors refer separately to the two frequency components; values below $10^{-14}$ are reported as limits.}
\label{tab:kerr_benchmark}
\papertablestyle
\begin{tabular}{rrrrr}
\tabletoprule
\multirow{2}{*}{$\chi$} & \multicolumn{2}{c}{Kerr frequency $\widehat\omega$} & \multicolumn{2}{c}{Relative difference} \\
\cmidrule[0.4pt](lr){2-3}\cmidrule[0.4pt](lr){4-5}
 & \multicolumn{1}{c}{Matching} & \multicolumn{1}{c}{Leaver} & \multicolumn{1}{c}{Real part} & \multicolumn{1}{c}{Imaginary part} \\
\midrule[0.45pt]
0.1 & $0.4994821995-0.0966660162i$ & $0.4994821995-0.0966660162i$ & $<10^{-14}$ & $<10^{-14}$\\
0.5 & $0.5859896293-0.0934944190i$ & $0.5859896293-0.0934944190i$ & $<10^{-14}$ & $<10^{-14}$\\
0.9 & $0.7816382887-0.0692893119i$ & $0.7816382887-0.0692893119i$ & $8.47\!\times\!10^{-13}$ & $4.05\!\times\!10^{-11}$\\
\tablebottomrule
\end{tabular}
\end{table}

For $m=2,1,-1$ we compute
\begin{equation}
C_h=\frac{\widehat\omega(h)-\widehat\omega(-h)}{2h},
\qquad
C^{\rm R}_h=\frac{4C_{h/2}-C_h}{3},
\label{eq:results_symmetric_response}
\end{equation}
using $h=0.02,0.01,0.005,0.0025$.
To see the advantage of this procedure, write a smooth local
expansion as $\widehat\omega(\eta)=\widehat\omega_{\rm K}
+C\eta+b_2\eta^2+b_3\eta^3+\cdots$. The symmetric difference cancels
the even powers and gives $C_h=C+b_3h^2+O(h^4)$; the Richardson
combination~\cite{RichardsonGaunt1927} then cancels the leading $h^2$ error. Decreasing $h$
indefinitely is not sufficient, however, because the difference
of two close frequencies magnifies their numerical uncertainties
by a factor proportional to $1/h$.
The reported derivative uses the last Richardson pair; we denote
this final estimate by $C_{\rm R}$. For $m=2,1,-1$ it is
$C_{\rm R}=(4C_{0.0025}-C_{0.005})/3$.
The responses labeled $C$ in the main root-difference results are
these extrapolated estimates, not a single finite-step value $C_h$. A
conservative uncertainty estimate combines its change from the
finest unextrapolated difference, the change under alternate
radial settings, and the root-correction contribution. The alternate
settings use 60-digit working precision, boundary-series order 20,
finite outer radius $r_\infty=50M$, horizon offset $\epsilon=10^{-6}M$,
and matching radius $r_{\rm match}=6M$.
Real and imaginary components are checked separately.
An imaginary response is labeled \emph{not numerically resolved} (NR)
when its sign changes across the tested steps or its uncertainty is
comparable to its magnitude. This criterion leaves the $m=2$,
$\chi=0.1$ imaginary response unresolved. The supplemental $m=-2$ scan
uses all four steps at $\chi=0.1,0.5,0.9$, including new fine-step roots
and alternate radial checks. Its other six spins retain the coarser pair
$h=0.02,0.01$, explicitly identified in the electronic table.
We also tabulate successive Richardson estimates to assess
derivative convergence.

\subsection{Independent radial and angular-truncation checks}
\label{sec:independent_checks}

The Kerr benchmark tests the undeformed operator but cannot alone validate
its deformation terms. We therefore solve the projected radial equation
independently by Chebyshev--Lobatto collocation~\cite{WeidemanReddy2000}.
With $x=1-r_+/r$, we write
\begin{equation}
 R=e^{i\omega r}x^{\alpha_H}r^{2iM\omega-1}y(x),
 \qquad \alpha_H=-\frac{i k_H(r_+^2+a^2)}{r_+-r_-}.
 \label{eq:spectral_ansatz}
\end{equation}
Here $R=X_{\ell m}/\sqrt{r^2+a^2}$ is the unrescaled radial amplitude,
and $k_H$ is the consistently chosen horizon wave number of the
truncated operator. Regular endpoint equations for $y$ replace the
finite-radius boundary series. A normalization of $y$ at an interior
node gives a bordered linear system; the unused differential-equation
row supplies the nonlinear frequency residual. We independently vary
$N=40,60,80,100$ at $\chi=0.1,0.5,0.9$ and $\eta=0,\pm0.01$.
The accepted $N=100$ roots agree with the matching roots within the
$10^{-12}$ reporting scale in both components
(Table~\ref{tab:independent_nonzero}). This provides an independent radial discretization of the same
projected equation.
\begin{table}[t]
\centering
\caption{Independent radial validation of the Method II truncated operator for $(2,2,0)$. The compactified Chebyshev calculation uses $N+1$ points and no finite-radius boundary series. $E_N$ is the maximum componentwise change between $N=80$ and $100$. Differences below $10^{-12}$ are displayed as limits.}
\label{tab:independent_nonzero}
\papertablestyle
\begin{tabular}{rrrrrr}
\tabletoprule
\multicolumn{2}{c}{Parameters} & \multicolumn{1}{c}{Spectral frequency} & \multicolumn{2}{c}{Difference from matching} & \multicolumn{1}{c}{Convergence} \\
\cmidrule[0.4pt](lr){1-2}\cmidrule[0.4pt](lr){4-5}
\multicolumn{1}{c}{$\chi$} & \multicolumn{1}{c}{$\eta$} & \multicolumn{1}{c}{$\widehat\omega_{\rm spectral}$} & \multicolumn{1}{c}{$|\Delta\operatorname{Re}\widehat\omega|$} & \multicolumn{1}{c}{$|\Delta\operatorname{Im}\widehat\omega|$} & \multicolumn{1}{c}{$E_N$} \\
\midrule[0.45pt]
0.1 & $-0.01$ & $0.4993170560-0.0966659967i$ & $<10^{-12}$ & $<10^{-12}$ & $<10^{-12}$ \\
0.1 & $+0.01$ & $0.4996475449-0.0966659968i$ & $<10^{-12}$ & $<10^{-12}$ & $<10^{-12}$ \\
\addlinespace[3pt]
0.5 & $-0.01$ & $0.5847142810-0.0934847908i$ & $<10^{-12}$ & $<10^{-12}$ & $<10^{-12}$ \\
0.5 & $+0.01$ & $0.5872739735-0.0935021728i$ & $<10^{-12}$ & $<10^{-12}$ & $<10^{-12}$ \\
\addlinespace[3pt]
0.9 & $-0.01$ & $0.7770076029-0.0692518709i$ & $<10^{-12}$ & $<10^{-12}$ & $<10^{-12}$ \\
0.9 & $+0.01$ & $0.7863501236-0.0692848181i$ & $<10^{-12}$ & $<10^{-12}$ & $<10^{-12}$ \\
\tablebottomrule
\end{tabular}
\end{table}

\begin{figure}[t]
\centering
\includegraphics[width=0.98\textwidth]{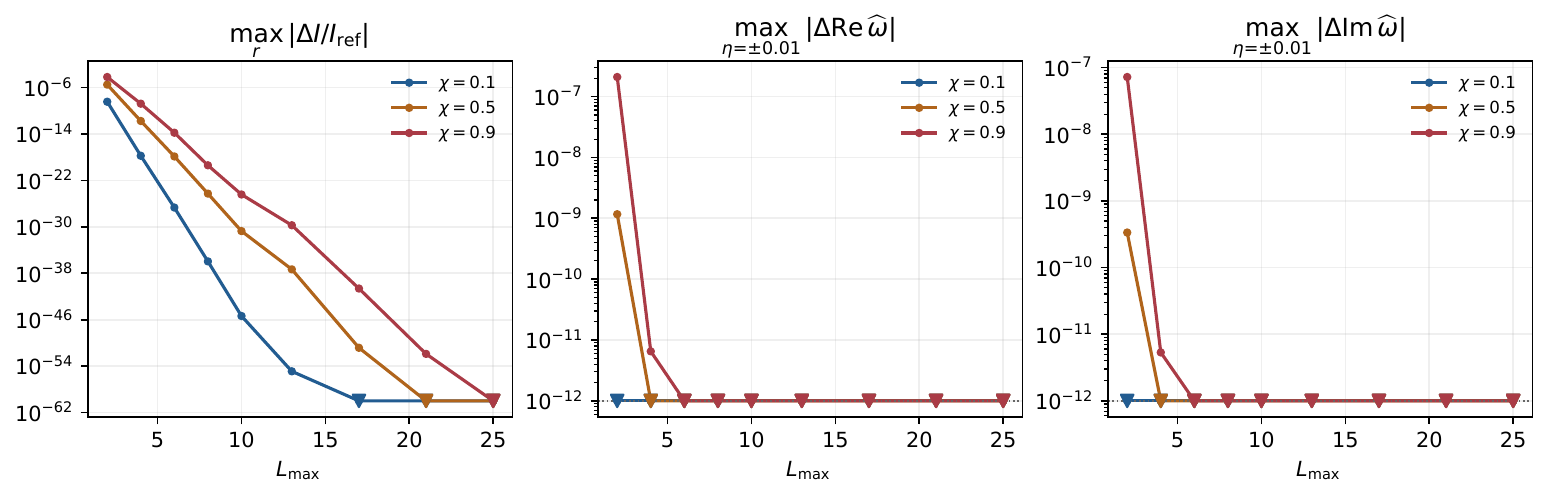}
\caption{Angular-basis truncation at fixed physical $(2,2,0)$.
Left: sampled relative projection error. Center and right: changes of
the recomputed frequency components, maximized over $\eta=\pm0.01$.
Downward triangles indicate differences below the reporting floors
($10^{-60}$ for the projection and $10^{-12}$ for the frequency). Lines only guide the eye
between calculated cutoffs.}
\label{fig:angular_qnm}
\end{figure}

For the angular test, the physical label $\ell=2$ is held fixed while
the expansion of each spheroidal eigenfunction in spherical harmonics
is truncated at $j\leq L_{\max}$. We vary
$L_{\max}=2,4,6,8,10,13,17,21,25$ at $N_G=96$ and compare with
$(L_{\max},N_G)=(25,128)$. The scalar angular matrix preserves parity,
so odd and even spherical channels are not mixed; an odd upper cutoff
need not add a nonzero coefficient to the $\ell=2,m=2$ branch.
We measure
\begin{equation}
 \epsilon_I(L_{\max})=
 \max_{r\in\mathcal R}
 \frac{|I_{222}^{(L_{\max},96)}(r)-I_{222}^{(25,128)}(r)|}
 {|I_{222}^{(25,128)}(r)|},
 \label{eq:angular_error_measure}
\end{equation}
on a specified 401-point grid $\mathcal R\subset[r_+,10M]$.
At every cutoff we then recompute the two $\eta=\pm0.01$ QNM roots
from displaced seeds, keeping the Kerr angular eigenvalue evaluation
unchanged. Thus this test isolates the projection's truncation error. We also recompute roots at $N_G=64,96,128$ with
$L_{\max}=25$. Table~\ref{tab:angular_qnm} and
Fig.~\ref{fig:angular_qnm} distinguish convergence of the integral
from its propagated frequency effect. For example, at $\chi=0.9$,
$L_{\max}=2$ gives $\epsilon_I=6.72\times10^{-5}$ and maximum
real/imaginary frequency changes $2.10\times10^{-7}$ and
$7.23\times10^{-8}$. At $L_{\max}=4$ these frequency changes
fall to $6.51\times10^{-12}$ and $5.34\times10^{-12}$;
for $L_{\max}\geq6$ they are below our reporting floor.
Thus the production cutoff $L_{\max}=17$ is sufficient
for these representative fundamental modes. Frequency changes below $10^{-12}$ are reported as upper limits.
\begin{table}[t]
\centering
\caption{Propagation of spherical-basis truncation into $I_{222}$ and the QNM frequency. The physical mode is fixed at $(2,2,0)$; $L_{\max}$ is the spherical-basis cutoff. Here $N_G=96$, and the reference is $(L_{\max},N_G)=(25,128)$. $\epsilon_I$ is the maximum relative difference on 401 radii from $r_+$ to $10M$; frequency changes are the maxima over independently recomputed $\eta=\pm0.01$ roots. Projection differences below $10^{-60}$ and frequency differences below $10^{-12}$ are reported as upper limits.}
\label{tab:angular_qnm}
\papertablestyle
\begin{tabular}{rrrrr}
\tabletoprule
\multirow{2}{*}{$\chi$} & \multirow{2}{*}{$L_{\max}$} & \multicolumn{1}{c}{Projection error} & \multicolumn{2}{c}{Frequency change} \\
\cmidrule[0.4pt](lr){4-5}
 &  & \multicolumn{1}{c}{$\epsilon_I$} & \multicolumn{1}{c}{$|\Delta\operatorname{Re}\widehat\omega|$} & \multicolumn{1}{c}{$|\Delta\operatorname{Im}\widehat\omega|$} \\
\midrule[0.45pt]
0.5 & 2 & $3.29\times10^{-6}$ & $1.17\times10^{-9}$ & $3.35\times10^{-10}$ \\
0.5 & 4 & $1.78\times10^{-12}$ & $<10^{-12}$ & $<10^{-12}$ \\
0.5 & 6 & $1.37\times10^{-18}$ & $<10^{-12}$ & $<10^{-12}$ \\
0.5 & 8 & $5.36\times10^{-25}$ & $<10^{-12}$ & $<10^{-12}$ \\
0.5 & 17 & $1.42\times10^{-51}$ & $<10^{-12}$ & $<10^{-12}$ \\
0.5 & 25 & $<10^{-60}$ & $<10^{-12}$ & $<10^{-12}$ \\
\addlinespace[3pt]
0.9 & 2 & $6.72\times10^{-5}$ & $2.10\times10^{-7}$ & $7.23\times10^{-8}$ \\
0.9 & 4 & $1.72\times10^{-9}$ & $6.51\times10^{-12}$ & $5.34\times10^{-12}$ \\
0.9 & 6 & $1.69\times10^{-14}$ & $<10^{-12}$ & $<10^{-12}$ \\
0.9 & 8 & $4.04\times10^{-20}$ & $<10^{-12}$ & $<10^{-12}$ \\
0.9 & 17 & $2.35\times10^{-41}$ & $<10^{-12}$ & $<10^{-12}$ \\
0.9 & 25 & $<10^{-60}$ & $<10^{-12}$ & $<10^{-12}$ \\
\tablebottomrule
\end{tabular}
\end{table}

Finally, we check the response extraction without subtracting two QNM
roots by differentiating the matching equation itself:
\begin{equation}
 C_{\rm imp}=-\left.
 \frac{\partial_\eta F}{\partial_{\widehat\omega}F}
 \right|_{\widehat\omega_{\rm K},0}.
 \label{eq:implicit_response}
\end{equation}
The trial-frequency derivative includes the Kerr angular eigenvalue,
and the deformation derivative includes the boundary wave number.
The projection remains frozen at the Kerr frequency, as required at
first order. Central differences with dimensionless steps
$10^{-4},5\times10^{-5},2.5\times10^{-5}$, Richardson extrapolation,
and alternate radial settings give agreement with the QNM-root
differences within their estimated uncertainties at all three
representative spins. This tests derivative extraction with the same
matching function.

\subsection{The $(2,2,0)$ mode and the Kerr reference}
\label{sec:results220}

\begin{figure}[t]
\centering
\includegraphics[width=0.95\textwidth]{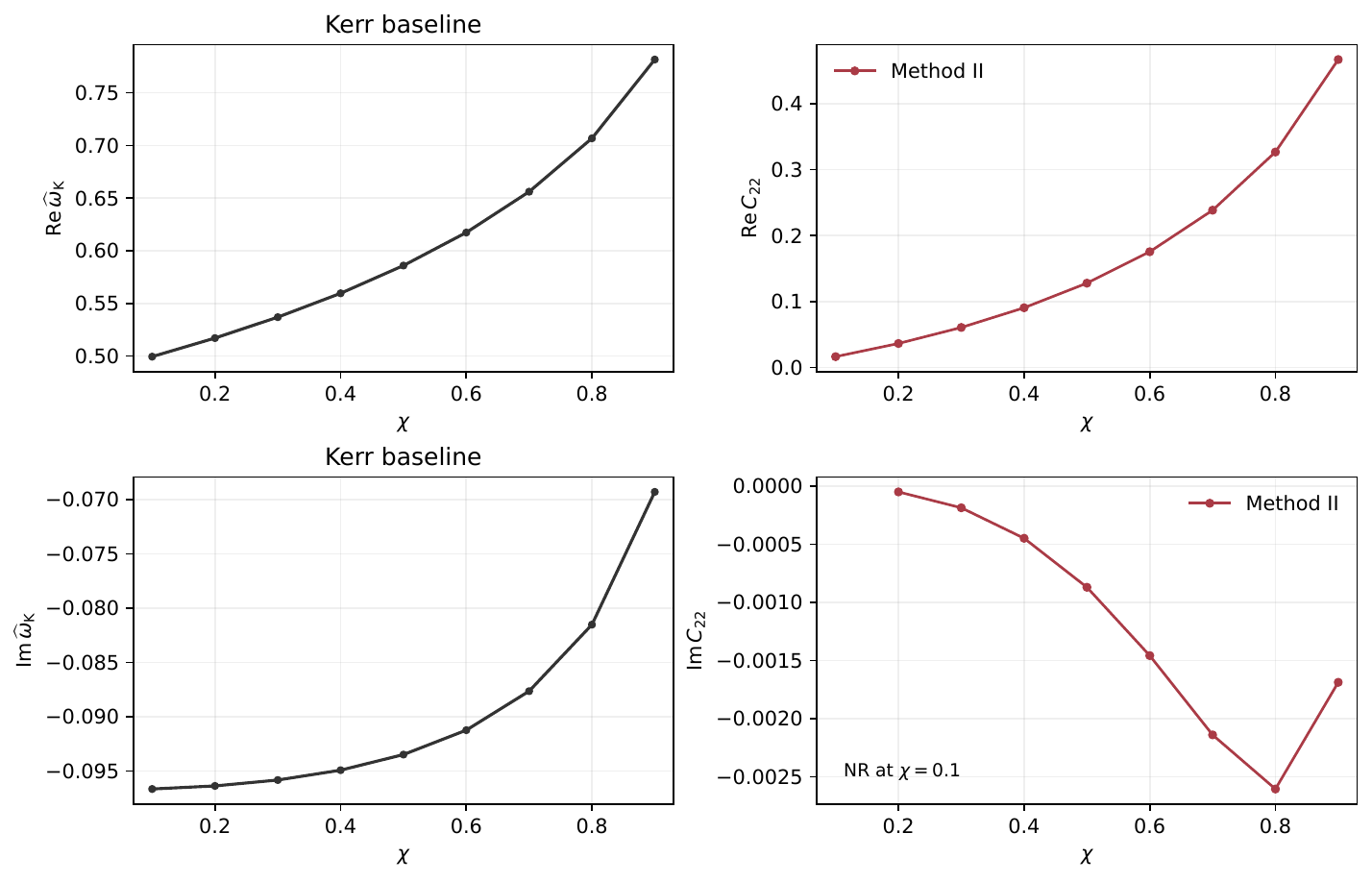}
\caption{Matched Kerr frequencies (left) and the Method II
linear response $C_{22}=\partial_\eta(M\omega)|_0$ (right).
Only computed grid points are shown. The NR imaginary response at $\chi=0.1$ is omitted. Numerical uncertainty estimates
are retained in the electronic data but are not drawn, since they
are smaller than the plotting symbols.}
\label{fig:scalar220}
\end{figure}

Figure~\ref{fig:scalar220} displays the Kerr frequency and the
linear Method II response. The Kerr real frequency increases
from $0.49948220$ at $\chi=0.1$ to $0.78163829$ at $\chi=0.9$,
while its imaginary part changes from $-0.09666602$ to
$-0.06928931$. The deformation response at fixed spin has
$\operatorname{Re}C_{22}=0.0165244$, $0.127983$, and $0.467086$
at $\chi=0.1,0.5,0.9$, respectively. A small positive deformation
therefore raises the oscillation frequency of this corotating
mode, with increasing sensitivity toward rapid rotation.

The resolved imaginary response is much smaller and is not
monotonic in spin. For example,
$\operatorname{Im}C_{22}=-0.00087009$ at $\chi=0.5$,
$-0.00260448$ at $\chi=0.8$, and $-0.00168711$ at $\chi=0.9$.
Since $\left.\partial_\eta\gamma\right|_{\eta=0}=-\operatorname{Im}C$, these values
correspond to increased damping for positive infinitesimal
$\eta$, even though increasing the Kerr spin reduces damping.

At leading order in spin, the frame-dragging terms imply
\begin{equation}
\widehat\omega_{\ell mn}
=\widehat\omega_{\ell n}^{\rm Sch}
+m\chi(1+\eta)c_{\ell n}+O(\chi^2).
\label{eq:results_leading_split}
\end{equation}
This explains the disappearance of the response as $a\to0$ and
the leading degeneracy between spin and deformation.

\subsection{Azimuthal dependence and the exact null test}

\begin{figure}[t]
\centering
\includegraphics[width=0.95\textwidth]{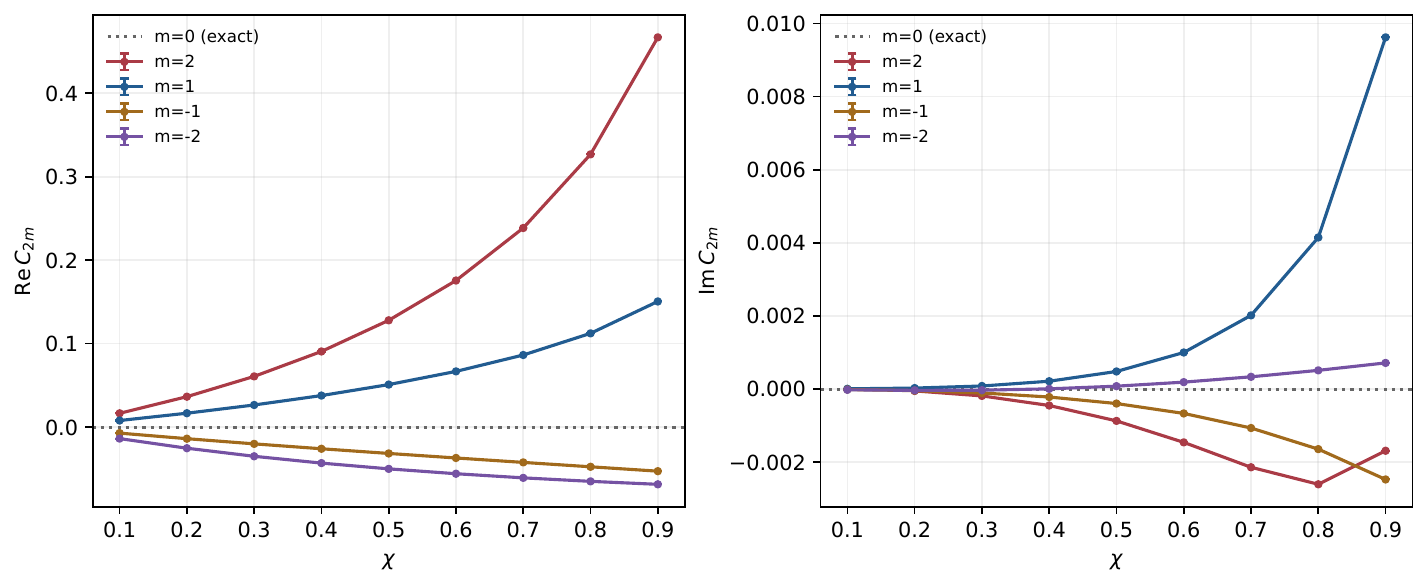}
\caption{Real and imaginary Method II linear responses for
$m=2,1,-1,-2$, with the exact $m=0$ null line. The $m=-2$
derivative uses four steps at $\chi=0.1,0.5,0.9$ and the coarser
pair $h=0.02,0.01$ elsewhere; all displayed
points pass residual and alternate-radial-setting checks.
NR or invalid-input points are omitted rather than interpolated.}
\label{fig:scalar_m}
\end{figure}

Figure~\ref{fig:scalar_m} compares the linear response of the
available azimuthal branches. The real response is positive for
$m=2,1$ and negative for $m=-1,-2$. 
At $\chi=0.9$,
$\operatorname{Re}C_{21}=0.150591$ and
$\operatorname{Re}C_{2,-1}=-0.0528007$, demonstrating that finite
spin breaks a simple antisymmetry. The distinct Kerr frequencies,
spheroidal eigenfunctions, and terms proportional to $m^2$
all enter this comparison.

The imaginary responses also differ qualitatively among the
branches: at $\chi=0.9$ they are approximately $0.0096222$
for $m=1$ and $-0.00247215$ for $m=-1$. Positive deformation
therefore need not change the damping in the same direction
for different modes. In contrast, $m=0$ is an exact null test. Both deformation terms in the
scalar equation vanish identically, and hence $C_{20}=0$ and the
axisymmetric scalar spectrum coincides with Kerr at fixed $M,a$.

\subsection{Comparison at common linear order}
\label{sec:results_sr}

\begin{figure}[t]
\centering
\includegraphics[width=0.95\textwidth]{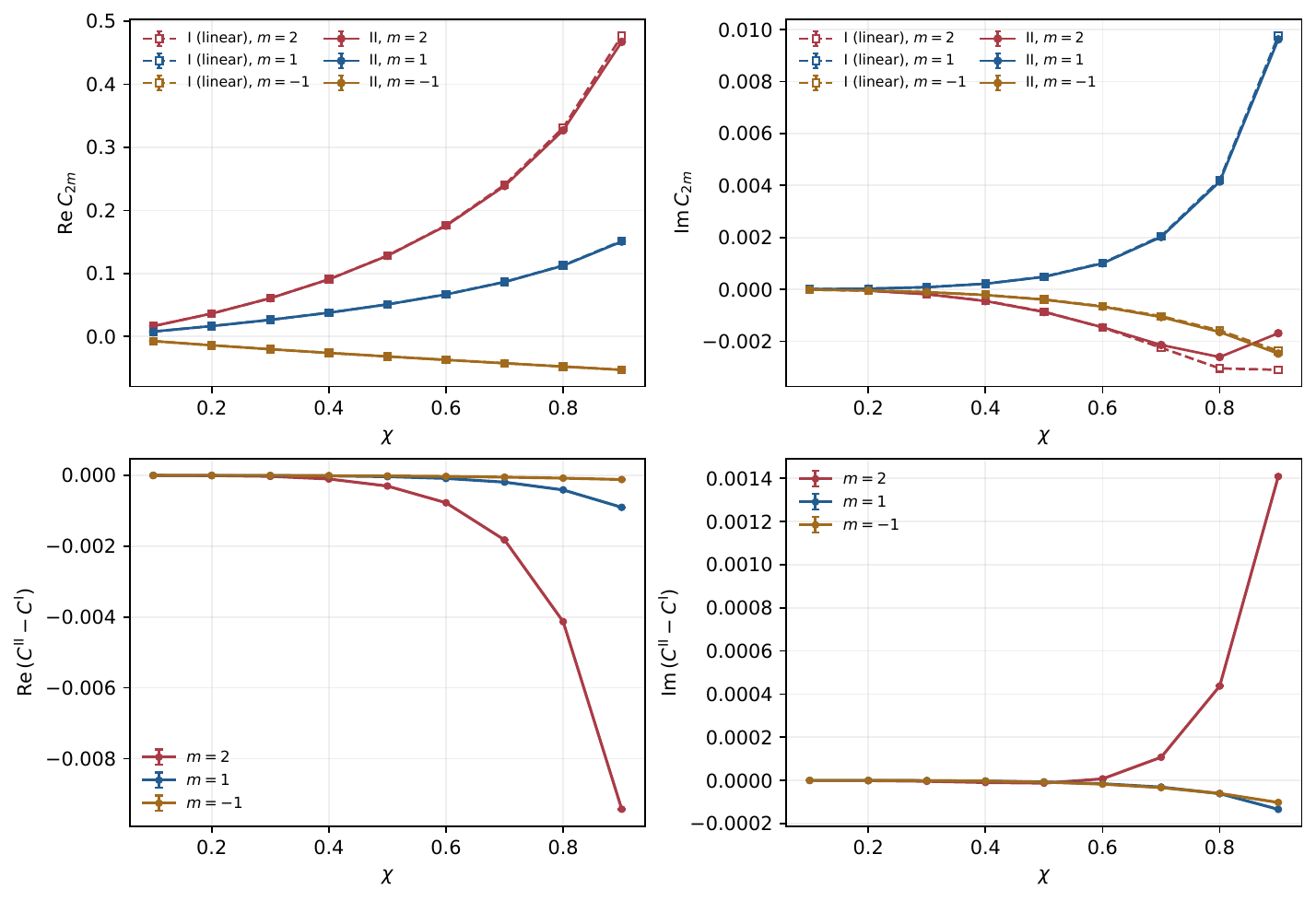}
\caption{Common-linear-order comparison for $\ell=2$, $n=0$,
and $m=2,1,-1$ in both rows. Upper panels show the responses from
linearized Method I (dashed lines, open squares) and projected
Method II (solid lines, filled circles), with the same color identifying
each $m$. Lower panels show $\Delta C$ for the same three branches.
Differences in real
and imaginary components are displayed separately; NR imaginary
entries are omitted.}
\label{fig:scalar_sr}
\end{figure}

To isolate the diagonal angular correction, Method I is also
expanded to $O(\eta)$ before computing its response $C^{\rm I}$.
Both calculations then use the same matching solver, numerical
settings, and finite-difference steps. The relevant difference is
\begin{equation}
\Delta C=C^{\rm II}-C^{\rm I}\rev{.}
\label{eq:results_discrepancies}
\end{equation} Figure~\ref{fig:scalar_sr} and
Table~\ref{tab:scalar220} show the comparison. Both rows of
Fig.~\ref{fig:scalar_sr} use the same branches $m=2,1,-1$ at
$\ell=2$, $n=0$: the upper row compares the two responses directly,
and the lower row isolates their difference. For $(2,2,0)$,
the relative difference in the real response grows from
$0.00140\%$ at $\chi=0.1$ to $0.234\%$ at $\chi=0.5$ and
$1.98\%$ at $\chi=0.9$. Thus the slow-rotation approximation
applied only to the nonseparable sector captures most of the real
linear response, while retaining the projected angular structure
becomes increasingly important at rapid rotation.

The imaginary response requires a separate assessment. At
$\chi=0.9$, Method I gives $\operatorname{Im}C^{\rm I}_{22}
=-0.00309592$, compared with $-0.00168711$ from Method II. The magnitude of their difference is approximately $45.5\%$ of the magnitude
of the Method I imaginary response.
A complex-frequency norm dominated by its real part would hide
this distinction. The inter-method difference measures the
retained diagonal angular contribution within their common
perturbative order.

\begin{table}[t]
\centering
\caption{Common-linear-order response $C=\partial_\eta(M\omega)|_0$ at fixed geometric $M,a$. NR denotes an unresolved imaginary response.}
\label{tab:scalar220}
\papertablestyle
\begin{tabular}{rrrrrr}
\tabletoprule
\multirow{2}{*}{$m$} & \multirow{2}{*}{$\chi$} & \multicolumn{2}{c}{Method I} & \multicolumn{2}{c}{Method II} \\
\cmidrule[0.4pt](lr){3-4}\cmidrule[0.4pt](lr){5-6}
 &  & \multicolumn{1}{c}{$\operatorname{Re}C^{\rm I}$} & \multicolumn{1}{c}{$\operatorname{Im}C^{\rm I}$} & \multicolumn{1}{c}{$\operatorname{Re}C^{\rm II}$} & \multicolumn{1}{c}{$\operatorname{Im}C^{\rm II}$} \\
\midrule[0.45pt]
2 & 0.1 & $0.0165246669$ & $5.23\times10^{-8}$ & $0.0165244360$ & NR \\
2 & 0.5 & $0.1282829$ & $-0.00085747$ & $0.1279829$ & $-0.00087009$ \\
2 & 0.9 & $0.476517$ & $-0.003096$ & $0.467086$ & $-0.001687$ \\
\addlinespace[3pt]
1 & 0.1 & $0.00789478483$ & $7.44163\times10^{-6}$ & $0.00789474935$ & $7.43075\times10^{-6}$ \\
1 & 0.5 & $0.05099897$ & $0.000488687$ & $0.05096392$ & $0.000481670$ \\
1 & 0.9 & $0.1514922$ & $0.00975617$ & $0.1505909$ & $0.00962220$ \\
\addlinespace[3pt]
$-1$ & 0.1 & $-0.00720564646$ & $-1.193376\times10^{-5}$ & $-0.00720567644$ & $-1.194498\times10^{-5}$ \\
$-1$ & 0.5 & $-0.031549500$ & $-0.000387542$ & $-0.031563761$ & $-0.000395486$ \\
$-1$ & 0.9 & $-0.05268529$ & $-0.00236999$ & $-0.05280065$ & $-0.00247215$ \\
\tablebottomrule
\end{tabular}
\end{table}

\subsection{Potential interpretation and finite-deformation diagnostics}

We now interpret the spectrum using the potential contributions
already defined in Secs.~\ref{sec:method1} and \ref{sec:small_LV}.
The frame-dragging term $\delta V_{\rm FD}$ is common to both methods.
At linear order, their distinction is the replacement of
$\delta V_{\rm sep}^{(1)}$ in Eq.~\eqref{eq:method1_sep_linear}
by $\delta V_{\rm proj}$ in Eq.~\eqref{eq:results_sep}, which retains
the diagonal angular projection at the Kerr reference frequency.
This comparison isolates the treatment of the nonseparable kernel
at $O(\eta)$.

\begin{figure}[t]
\centering
\includegraphics[width=0.95\textwidth]{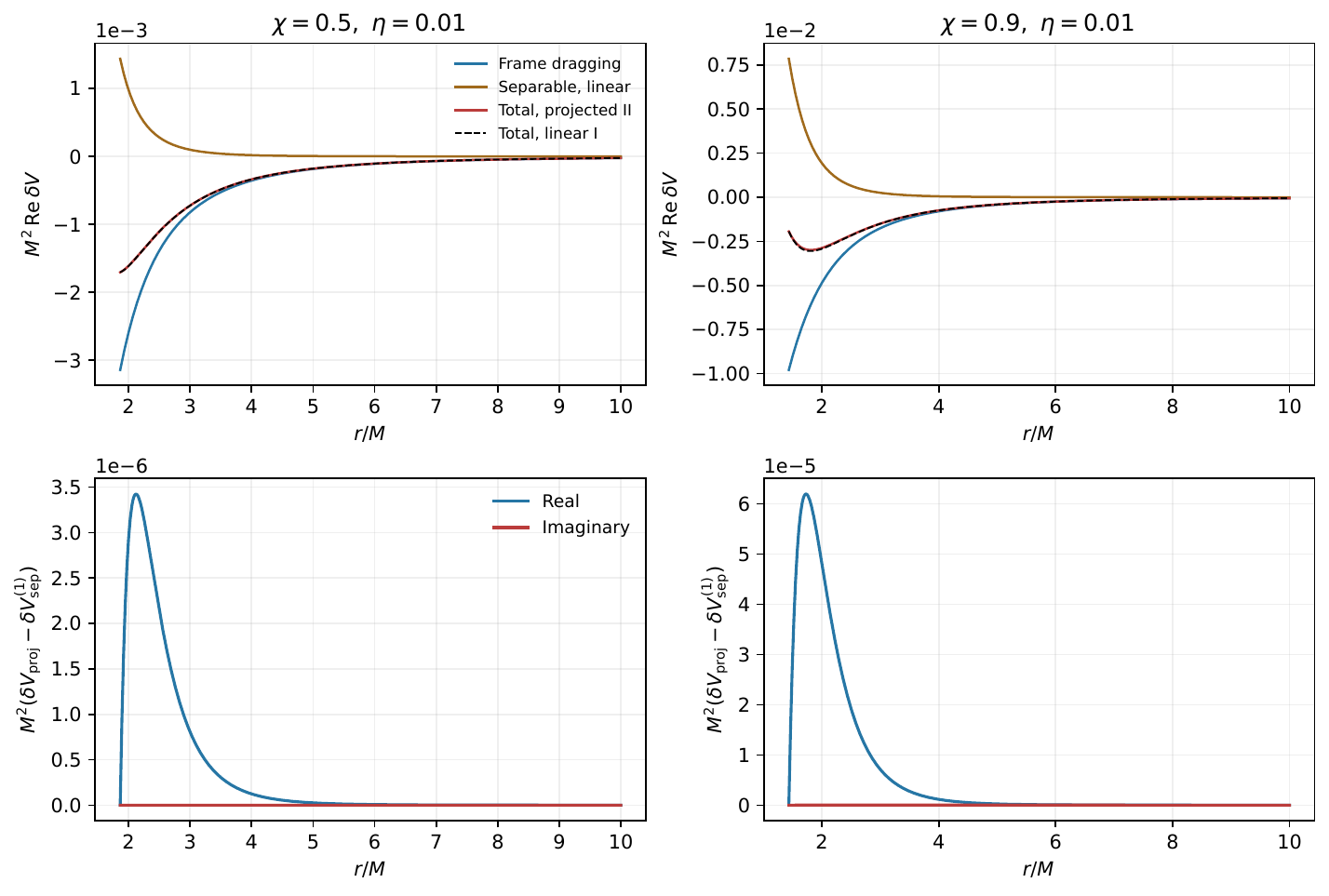}
\caption{Common-order potential corrections at $\eta=0.01$
and the matched Kerr frequency, for $(2,2,0)$. Upper panels
separate frame dragging, the linear separable term, and the total
with or without projection. Lower panels isolate the projected
angular contribution. No quadratic deformation term is included
in this common-order comparison.}
\label{fig:scalar_potential}
\end{figure}

Figure~\ref{fig:scalar_potential} compares these terms at a shared
Kerr frequency. The negative real frame-dragging term partially
cancels the positive separable term for a corotating mode with
$\eta>0$. The projection adds both a real and a small imaginary
correction. It agrees with the separable factor at the horizon,
where $I_{\ell\ell m}(r_+)=D_+^{-2}$ and $D_+=r_+^2+a^2$. These are complex,
frequency-dependent wave coefficients, so their local signs or
amplitudes do not directly determine a QNM-frequency shift.

\begin{figure}[t]
\centering
\includegraphics[width=0.95\textwidth]{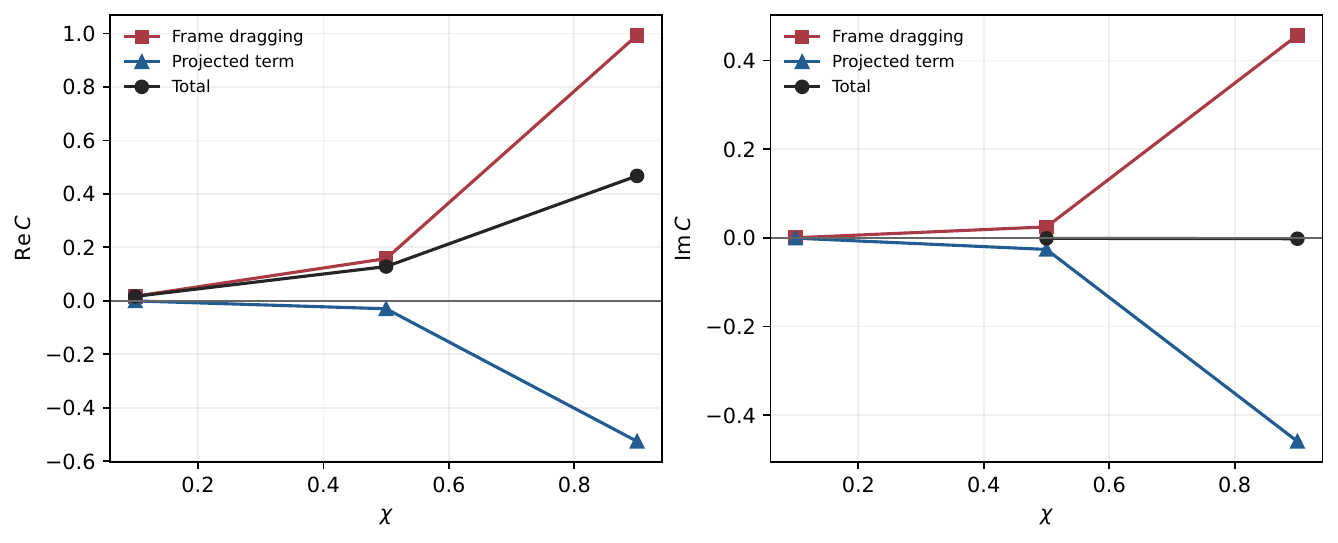}
\caption{Frequency-response contributions obtained from
Eq.~\eqref{eq:implicit_response}. The projected contribution competes
with frame dragging; their cancellation is especially pronounced in
the imaginary component. The NR total imaginary response at
$\chi=0.1$ is not plotted. Only the three calculated spins are shown.}
\label{fig:response_decomposition}
\end{figure}

To make this distinction quantitative, we separately differentiate
the matching residual with only the frame-dragging or projected term
activated, including the corresponding first-order boundary change.
Linearity predicts $C=C_{\rm FD}+C_{\rm proj}$; we verify this against
an independently evaluated derivative of the complete operator
(Table~\ref{tab:response_decomposition}).
Figure~\ref{fig:response_decomposition} displays these frequency-response
contributions: the left panel shows the real part and the right panel
the imaginary part. In the real part, the positive frame-dragging
response is partially offset by the negative projected response,
leaving a positive total. The imaginary contributions almost cancel,
so the net damping response is much smaller than either contribution.
At $\chi=0.9$ the imaginary
contributions are approximately $+0.457102$ and $-0.458789$, leaving
a total of only $-0.00168711$. Consequently, a small change in one
contribution can be a substantial fraction of the total damping
response while remaining a small correction to the real response.
\begin{table}[t]
\centering
\caption{First-order response decomposition obtained by differentiating the matching residual at the Kerr root. The two operator terms include their consistent horizon-boundary variations. The sum is tested against a separate derivative of the complete projected operator, rather than imposed.}
\label{tab:response_decomposition}
\papertablestyle
\begin{tabular}{rrrrrrr}
\tabletoprule
\multirow{2}{*}{$\chi$} & \multicolumn{3}{c}{Real response} & \multicolumn{3}{c}{Imaginary response} \\
\cmidrule[0.4pt](lr){2-4}\cmidrule[0.4pt](lr){5-7}
 & \multicolumn{1}{c}{$\operatorname{Re}C_{\rm FD}$} & \multicolumn{1}{c}{$\operatorname{Re}C_{\rm proj}$} & \multicolumn{1}{c}{$\operatorname{Re}C$} & \multicolumn{1}{c}{$\operatorname{Im}C_{\rm FD}$} & \multicolumn{1}{c}{$\operatorname{Im}C_{\rm proj}$} & \multicolumn{1}{c}{$\operatorname{Im}C$} \\
\midrule[0.45pt]
0.1 & $1.70\times10^{-2}$ & $-4.83\times10^{-4}$ & $1.65\times10^{-2}$ & $4.35\times10^{-4}$ & $-4.35\times10^{-4}$ & NR \\
0.5 & $1.58\times10^{-1}$ & $-2.99\times10^{-2}$ & $1.28\times10^{-1}$ & $2.50\times10^{-2}$ & $-2.58\times10^{-2}$ & $-8.70\times10^{-4}$ \\
0.9 & $9.93\times10^{-1}$ & $-5.26\times10^{-1}$ & $4.67\times10^{-1}$ & $4.57\times10^{-1}$ & $-4.59\times10^{-1}$ & $-1.69\times10^{-3}$ \\
\tablebottomrule
\end{tabular}
\end{table}

\begin{figure}[t]
\centering
\includegraphics[width=0.95\textwidth]{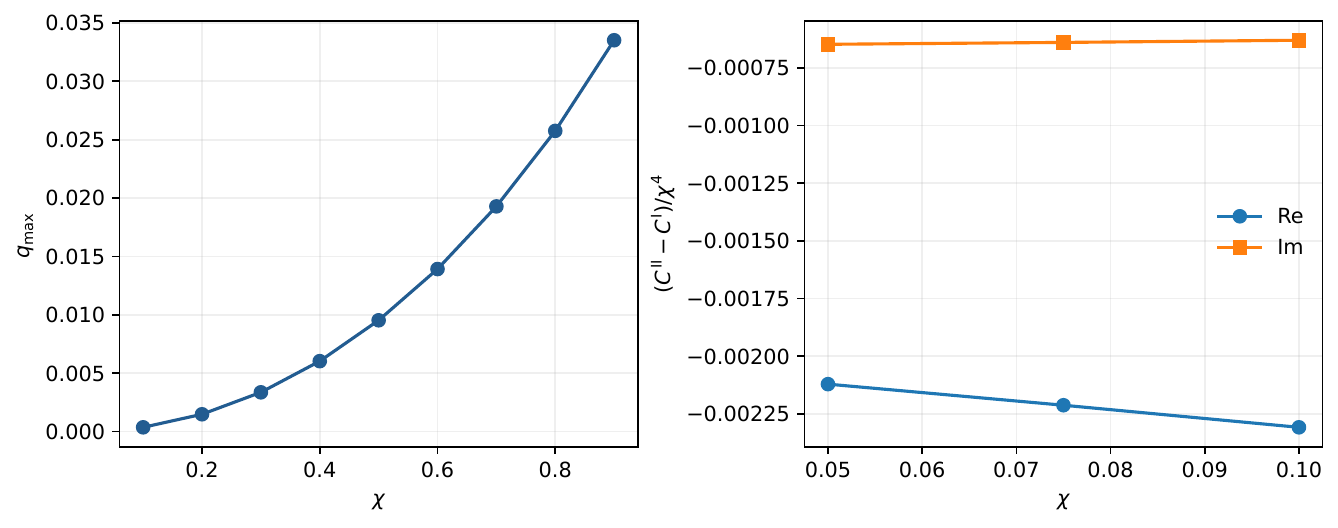}
\caption{Left: the geometric kernel-expansion parameter maximized
outside the horizon. Right: directly computed common-order angular
response differences divided by $\chi^4$ at three additional
low-spin settings.}
\label{fig:geometric_control}
\end{figure}

The left panel of Fig.~\ref{fig:geometric_control} shows
$q_{\rm max}(\chi)$, the maximum exterior value of the kernel-expansion
parameter defined in Sec.~\ref{sec:method1}. It reaches only
$q_{\rm max}=0.0335158$ at $\chi=0.9$, with its maximum at
$r/M=2.56000$. It controls the kernel expansion. The right panel provides a separate
low-spin check by plotting the real and imaginary parts of
$\Delta C/\chi^4$ for $(2,2,0)$. Direct response derivatives at
$\chi=0.05,0.075,0.1$ give
$\operatorname{Re}\Delta C/\chi^4=-0.0021212,-0.0022126,-0.0023083$.
The corresponding imaginary ratios are
$-0.0006482,-0.0006401,-0.0006306$.
Their modest variation is consistent with the leading $O(\eta a^4)$
omitted angular operator. Componentwise derivative
uncertainties and the sampled ratios are supplied electronically.

For finite deformation, Method I retains the full algebraic
$\eta(\eta+2)$ dependence of $\delta V_{\rm sep}$ in
Eq.~\eqref{eq:results_fd}, whereas Method II omits the original
operator's $O(\eta^2)$ terms. Their explicit difference at a common
frequency is
\begin{equation}
\delta V_{\rm II}-\delta V_{\rm I}
=\frac{8\eta a^2M^2r^2m^2}{D^2}(I_{\ell\ell m}-D^{-2})
-\frac{4\eta^2a^2M^2r^2m^2}{D^4}.
\label{eq:results_operator_difference}
\end{equation}
Moreover, the Method II horizon coefficient equals
$[\omega-m(1+\eta)\Omega_H^{\rm K}]^2-
\eta^2m^2(\Omega_H^{\rm K})^2$. Its ingoing square-root branch
therefore agrees with the exact horizon prescription only to
the retained order. Nonlinear eigenfrequency dependence produced
by solving this truncated operator does not restore its omitted
terms.

Continuation with 120-digit working precision gives, at
$(\chi,\eta)=(0.9,0.1)$, the converged Method II diagnostic
$\widehat\omega=0.8328292645-0.0668707011i$, with
$|F|<2.0\times10^{-18}$. An independent radial setting checks
this result. Table~\ref{tab:finite_diagnostic} compares the
recomputed $\eta=0.05,0.1$ diagnostics with Method I retaining
its full algebraic deformation factor. These finite-$\eta$ results serve as diagnostics of the truncated
operators; the physical comparison uses the small-$\eta$ linear response.

\begin{table}[t]
\centering
\caption{Finite-$\eta$ diagnostics at $\chi=0.9$, $(2,2,0)$, recomputed with 120-digit working precision and checked with an alternate radial setting. Method I retains its algebraic $\eta(\eta+2)$ term; Method II solves the first-order truncated operator.}
\label{tab:finite_diagnostic}
\papertablestyle
\begin{tabular}{rrrrr}
\tabletoprule
\multirow{2}{*}{$\eta$} & \multicolumn{2}{c}{Method I} & \multicolumn{2}{c}{Method II} \\
\cmidrule[0.4pt](lr){2-3}\cmidrule[0.4pt](lr){4-5}
 & \multicolumn{1}{c}{$\widehat\omega_{\rm I}$} & \multicolumn{1}{c}{$|F_{\rm I}|$} & \multicolumn{1}{c}{$\widehat\omega_{\rm II}$} & \multicolumn{1}{c}{$|F_{\rm II}|$} \\
\midrule[0.45pt]
0.05 & $0.8058294409-0.0694967881i$ & $5.88\!\times\!10^{-19}$ & $0.8060587919-0.0687945892i$ & $1.79\!\times\!10^{-18}$\\
0.10 & $0.8307071983-0.0697879332i$ & $3.67\!\times\!10^{-21}$ & $0.8328292645-0.0668707011i$ & $1.97\!\times\!10^{-18}$\\
\tablebottomrule
\end{tabular}
\end{table}

\section{Conclusions}\label{conclusion}

We have studied the fundamental quasinormal spectrum of a minimally
coupled massless scalar on a rotating Lorentz-violating black-hole
background. The exact scalar equation is nonseparable, although its
Kerr limit separates. Our two approximations use this limit without
expanding the Kerr radial or angular operators. Method I applies a
slow-rotation approximation only to the nonseparable deformation
kernel, retaining the already separable frame-dragging term and the
algebraic dependence on $\eta$. Method II instead expands in $\eta$
and keeps the diagonal biorthogonal spheroidal projection of the
angle-dependent correction. Comparing their frequency derivatives
at $\eta=0$ isolates the effect of this angular structure at a common
perturbative order.

The principal physical result is the complex linear response at fixed
geometric mass and spin. For the corotating branches studied, a small
positive deformation raises the real frequency, with a response that
increases with spin; the counterrotating branches have negative real
responses. The damping response is not governed by the same simple
trend. For $(2,2,0)$ its resolved values are negative and nonmonotonic
in spin, corresponding to increased damping for positive infinitesimal
$\eta$. Other branches can respond differently: at $\chi=0.9$, the
$m=1$ mode becomes less strongly damped, whereas the $m=-1$ mode
becomes more strongly damped. For $m=0$, both deformation terms vanish
identically, so the scalar spectrum is exactly Kerr at fixed $M,a$.

The two methods agree closely in the real linear response at small
spin. At $\chi=0.9$, their $(2,2,0)$ real responses differ by about
$2\%$, while their imaginary responses are $-0.00309592$ and
$-0.00168711$ for Methods I and II, respectively. The latter difference
has a magnitude of about $45.5\%$ of the magnitude of the Method I imaginary response. The separate response contributions explain this
contrast: frame dragging and angular projection nearly cancel in
the imaginary part, leaving a small net response that is relatively
sensitive to angular structure. Consequently, a good approximation
to the real shift does not guarantee comparable relative accuracy
for the damping shift. 

The numerical analysis combines residual-controlled continuation,
radial and angular convergence tests, Richardson-extrapolated
deformation derivatives, an independent Kerr continued-fraction
benchmark, and a nonzero-deformation radial spectral check. Angular
truncation tests follow both the projection integral and its effect
on the final frequency. Extending the physical
comparison beyond linear order requires the omitted deformation
terms and the appropriate coupled angular dynamics. Resolving the
remaining small damping response and extending the analysis to
additional multipoles and overtones are further useful steps.

\section*{Data availability}
The numerical source, raw root diagnostics, convergence tables,
and figure-generation scripts are included in the accompanying
numerical reproducibility package.

\section{Acknowledgments}

AI-assisted tools were used for code development, numerical diagnostics,
and manuscript preparation. Responsibility for the scientific content
and its validation remains with the authors.

\begingroup\looseness=-1

This work was supported by the National Key Research and Development Program of China (Grant No. 2023YFC2206701), the National Natural Science Foundation of China (Grant Nos. 12475056, 12605108, 12205129,  and 12247101), the 111 Project (Grant No. B20063),  Natural Science Foundation of Gansu Province (No. 22JR5RA389, No. 25JRRA799), the Gansu Province's Top Leading Talent Support Plane. Wen-Di Guo was supported by ``Talent Scientific Fund of Lanzhou University''.
\par\endgroup

%

%
%

\end{document}